\documentclass[preprint,10pt,5p,twocolumn,times]{elsarticle}
 \pdfoutput=1
\journal{Journal of Quantitative Spectroscopy and Radiative Transfer}

\usepackage{hyperref}
\usepackage{graphicx}
\usepackage[usenames,dvipsnames]{xcolor}
\usepackage{epstopdf}
\usepackage{bm}
\usepackage{dcolumn}
    \newcolumntype{d}[1]{D{.}{.}{#1}}
\usepackage{pbox}
\usepackage{amsmath}
\usepackage[section]{placeins}
\usepackage{threeparttable}
\usepackage[numbers]{natbib}

\newcommand{\wavenumbers}{cm$\rm{^{-1}}$}

\newcommand{\Avv}{Einstein  $A_{\mathrm{v}^\prime \mathrm{v}}$}
\newcommand{\fvv}{f_{\mathrm{v}^\prime \mathrm{v}}}
\newcommand{\AJJ}{A_{J\primed F\primed J F}}

\newcommand{\LEVEL}{\textsc{level}}
\newcommand{\PGO}{\textsc{pgopher}}
\newcommand{\RKR}{\textsc{rkr1}}
\newcommand{\MOLPRO}{\textsc{molpro}}

\newcommand{\etal}{\textit{et al.}}
\newcommand{\Dprimed}{^{\prime\prime}}
\newcommand{\primed}{^\prime}

\newcommand{\expo}[1]{\times 10^{#1}}

\newcommand{\DP}{$^2\Pi$}

\newcommand{\DSp}{$^2\Sigma^+$}
\newcommand{\TSm}{$^3\Sigma^-$}

\newcommand{\specialcell}[3][c]{
  \hspace{-1mm}\begin{tabular}[#1]{@{}#2@{}}#3\end{tabular}
  }

\renewcommand{\AJJ}{A_{J\primed J}}

\renewcommand{\LEVEL}{{\large\textsc{level}}}
\renewcommand{\RKR}{{\large\textsc{rkr1}}}
\renewcommand{\PGO}{{\large\textsc{pgopher}}}
\renewcommand{\MOLPRO}{{\large\textsc{molpro}}}

\begin{document}
\begin{frontmatter}

\title{Line Strengths of Rovibrational and Rotational Transitions in the X\DP\ Ground State of OH} 

\author{James S. A. Brooke\corref{cor1}\fnref{fn1}}
\address{Department of Chemistry, University of York, York, YO10 5DD,
UK; jsabrooke@gmail.com; +44 1904 434525.  }
\cortext[cor1]{Corresponding Author}
\fntext[fn1]{Now at School of Chemistry, University of Leeds, Leeds, LS2 9JT, UK. }

\author{Peter F. Bernath}
\address{Department of Chemistry \& Biochemistry, Old Dominion University, 4541 Hampton Boulevard, Norfolk, VA 23529-0126, USA; and Department of Chemistry, University of York, York, YO10 5DD,
UK.}

\author{Colin M. Western}
\address{School of Chemistry, University of Bristol, Cantock's Close, Bristol, BS8 1TS, UK.}

\author{Christopher Sneden}
\address{ Department of Astronomy, University of Texas at Austin, Austin, TX 78712, USA.}

\author{Melike Af\c{s}ar}
\address{Department of Astronomy and Space Sciences, Ege University,
35100 Bornova, \.{I}zmir, Turkey}

\author{Gang Li\fnref{fn2}}
\author{Iouli E. Gordon}
\address{Harvard-Smithsonian Center for Astrophysics, Atomic and Molecular Physics Division, Cambridge, MA 02138, USA.}
\fntext[fn2]{Now at Physikalisch-Technische Bundesanstalt (PTB), Bundesallee 100,  38116 Braunschweig, Germany}

\begin{abstract}
A new line list including positions and absolute intensities (in the form of Einstein $A$ values and oscillator strengths) has been produced for the OH ground X\DP\ state rovibrational (Meinel system) and pure rotational transitions. All possible transitions are included with v$\primed$ and v$\Dprimed$ up to 13, and $J$ up to between 9.5 and 59.5, depending on the band. An updated fit to determine molecular constants has been performed, which includes some new rotational data and a simultaneous fitting of all molecular constants. The absolute line intensities are based on a new dipole moment function, which is a combination of two high level ab initio calculations. The calculations show good agreement with an experimental v=1 lifetime, experimental $\mu_\mathrm{v}$ values, and $\Delta$v=2 line intensity ratios from an observed spectrum. To achieve this good agreement, an alteration in the method of converting matrix elements from Hund's case (b) to (a) was made. Partitions sums have been calculated using the new energy levels, for the temperature range 5-6000 K, which extends the previously available (in HITRAN) 70-3000 K range. The resulting absolute intensities have been used to calculate O abundances in the Sun, Arcturus, and two red giants in the Galactic open and globular clusters M67 and M71. Literature data based mainly on [O I] lines are available for the Sun and Arcturus, and excellent agreement is found.

\end{abstract}


\end{frontmatter}

\section{Introduction}\label{sec:Intro}
The OH radical is very important in atmospheric chemistry due to its high reactivity with volatile organic compounds~\citep{2003Atkinson-a,2004Lelieveld-a}, and it is the major oxidizing species in the lower atmosphere~\citep{1995Prinn-a}. OH is also produced in the upper atmosphere in excited vibrational levels, and near infrared emission to lower levels is the main cause of the nighttime airglow of the upper atmosphere~\citep{1950Meinel-a,1992Oliva-a}. This airglow is often an unwanted feature in astronomical observations~\citep{1993Maihara-a}, but has sometimes been used for wavelength calibration~\citep{1992Oliva-a} and atmospheric temperature retrievals~\citep{1987Sivjee-a}. OH is also present in stars~\citep{1972Wilson-a}, interstellar clouds~\citep{1968Heiles-a}, extra-terrestrial atmospheres~\citep{2008Piccioni-a,1994Atreya-a}, and often in relatively large concentrations~\citep{2003Settersten-a} in flames~\citep{1976Maillard-a, 1986Ewart-a, 1994Abrams-a}.

The transitions of interest in this work are in the Meinel system, which are the rovibrational transitions within the X\DP\ ground state. These have been used to determine the oxygen abundance in the Sun~\citep{1984Grevesse-a} and other stars, for example by \citet{2002Melendez-a} and \citet{2013Smith-a}.

\begin{figure}
  \includegraphics[width=9cm]{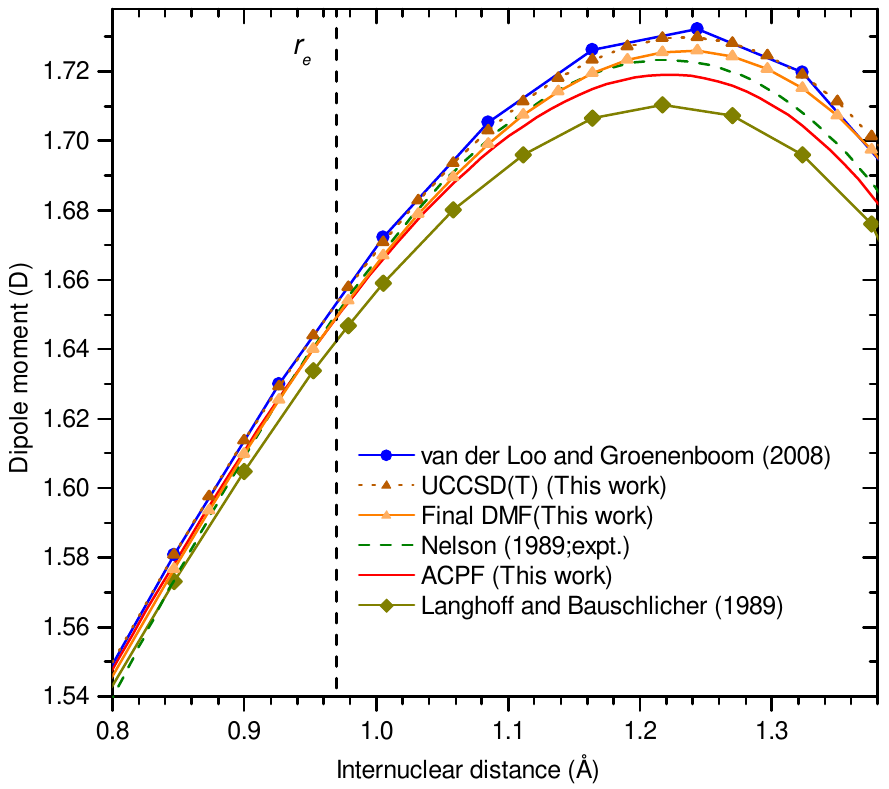}
  \centering
  \caption{Calculated and experimental dipole moment functions for the OH X\DP\ state.} \label{fig:DMFs}
\end{figure}

There have been several ab initio dipole moment functions (DMFs) calculated for the OH ground state~\citep{1974Stevens-a,1983Werner-a,1986Langhoff-a,1989Langhoff-a,2007vanderLoo-a,2008vanderLoo-a}. Accurate experimental dipole moments were obtained for the v=0, 1, and 2 levels by \citet{1984Peterson-a}, and these were used by \citeauthor{1990Nelson-a} \citep{1989Nelson-a,1990Nelson-a} in combination with their own experimental relative line intensities to calculate a DMF at internuclear distances between 0.70 and 1.76 $\mathrm{\AA}$. Another experimental DMF was obtained by \citet{1988Turnbull}, also using the \citeauthor{1984Peterson-a} values and their own experimental intensities. The various DMFs have all disagreed to some extent~\citep{1990Nelson-a} (also see Figure \ref{fig:DMFs}), including around the equilibrium internuclear distance ($r_e$), where the intensities are most sensitive to the DMF.

A set of transition probabilities was calculated by \citet{1974Mies-a}, and more recently \citet{1998Goldman-a} produced a set of intensities, which is now the most widely used for abundance calculations, for example by \citet{2002Melendez-a}. This list by \citeauthor{1998Goldman-a} is currently in the HITRAN~\citep{2013Rothman-a} database, and more extensively in the HITEMP~\citep{2010Rothman-a} database . The intensities are based on the DMF of \citet{1990Nelson-a} between 0.70 and 1.76 $\mathrm{\AA}$, and that of \citet{1989Langhoff-a} at other internuclear distances. Another line list including ab initio Einstein $\AJJ$ values was calculated recently~\citep{2007vanderLoo-a,2008vanderLoo-a}, however the accuracy can be improved, and it did not cover all of the available vibrational levels (up to v=9 as opposed to v=13). This DMF is included in the comparisons in Figure \ref{fig:DMFs}.

\citet{2009Bernath-b} performed a fit to obtain molecular constants up to v=13 in 2009, using the best data available at the time. Lines from the extensive study of \citet{1995Melen-a} were included, along with the rovibrational measurements of \citet{1994Abrams-a} (22 bands, $\Delta \mathrm{v}$=1-3, up to v=10), \citet{2001Nizkorodov-a} (11-9 band), and \citet{1990Sappey-a} (12-8 band), and B\DSp-X\DP\ system lines from \citet{1993Copeland-a} (v=7-13 in the X\DP\ state and v=0-1 in the B\DSp\ state).  They also assigned new rotational lines in v=4 in an ATMOS (1994)~\citep{1996Abrams-a} solar spectrum.

The fit of \citet{1995Melen-a} included rotational lines in v=0-3, from both their assignments of an ATMOS (1985)~\citep{1989Farmer-Book-a} solar spectrum and laboratory measurements, and rovibrational laboratory data from \citet{1976Maillard-a}. Hyperfine rotational data in v=0-3 from several other studies~\citep{1991Hardwick-a,1993Varberg-a,1986Brown-a,1979Davies-a,1968Radford-a,1970Ball-a,1975Meerts-a,1979Coxon-a,1974Destombes-a,1975Destombes-a,1981Kolbe-a,1972terMeulen-a,1976terMeulen-b, 1975Destombes-b,1976terMeulen-a,1979Meerts-a,1979Destombes-a,1971Clough-a,1971Lee-a} were also included in the fit by \citet{1995Melen-a}. Despite hyperfine structure not being included in the results of \citet{2009Bernath-b}, they took some of this information into account by adding pseudo-transitions to the fit, between $e$ and $f$ parity levels and various $J$ levels (``$\Lambda$-doubling data"), based on the term values calculated by \citet{1995Melen-a}. For a further description of the sources of the included transitions, the reader is referred to Table 1 of \citet{2009Bernath-b} and Figure 4 of \citet{1995Melen-a}.

Due to limitations in the fitting program used by \citet{2009Bernath-b}, four separate fits were required to calculate all of the molecular constants. In addition, more rotational data in v=0-2 has recently been reported~\citep{2011Martin-Drumel-a}, which has a higher reported accuracy than the previously available lines. A fit in which this new data was included and all parameters were floated simultaneously was performed in this work, and is described in Section \ref{sec:MolecularConstantFit}.

In this paper, the currently available HITRAN intensities are compared with experimental data (a v=1 lifetime~\citep{2005Meerakker-a}, $\Delta$v=2 H-W ratios (this work), and $\mu_\mathrm{v}$ values \citep{1984Peterson-a}). Based on these comparisons, it was concluded that the calculation of a new set of line intensities would be beneficial, and Sections \ref{sec:Spectrum} to \ref{sec:LineList} describe their calculation and validation. Partition function calculations are described in Section \ref{sec:Partition}, and Section \ref{ohabunds} uses the newly calculated line intensities to calculate the O abundance in the Sun and three other stars, to demonstrate their use in astronomical calculations.
\section{Molecular Constant Fit}\label{sec:MolecularConstantFit}
A new fit of molecular constants was performed, which was based mainly on that of \citet{2009Bernath-b}. Line weightings from the previous fit were retained, all parameters were floated simultaneously, and lines were replaced by new pure rotational data from \citet{2011Martin-Drumel-a} where possible, which was the only new data included. The $\Lambda$-doubling data included by \citet{2009Bernath-b} was also included here, and as with their fit, no hyperfine structure constants were calculated.

The fit of \citet{2009Bernath-b} was very extensive in terms of the number of molecular constants obtained. A number of higher order constants that had unreasonable fitting uncertainties were fixed at values based on extrapolation and careful inspection of the energy level patterns.

The same molecular constants as used by \citet{2009Bernath-b} were used here, and those that were fixed previously were also fixed. Floating these constants was attempted, but their resulting uncertainties were too large. As all parameters were now being floated, the uncertainties in general increased, and some of the higher order constants that were successfully determined in the fit of \citet{2009Bernath-b} were fixed at their previous values. Compared to the work of \citet{2009Bernath-b}, the small amount of new data and simultaneous determination of all constants resulted in only a small improvement in the fit. Specifically, the total unweighted average line position error changed from 0.02649 to 0.02637 \wavenumbers, and the weighted average error improved by a factor of 1.32.

See Table \ref{tab:MolecularConstants} for the final molecular constants.

\begin{table*}
\centering
\resizebox{15cm}{!} {
  \begin{threeparttable}
     \caption{\label{tab:MolecularConstants} Molecular constants$^\mathrm{a}$  for the OH X\DP\ state (in \wavenumbers). }
     \smallskip
     \begin{footnotesize}
        \begin{tabular}{@{} D{.}{.}{1}  D{.}{.}{13} D{.}{.}{12} D{.}{.}{14} D{.}{.}{11} D{.}{.}{8} D{.}{.}{8}}
        \hline\noalign{\smallskip}
        \mathrm{v} & \multicolumn{1}{c}{$T$} & \multicolumn{1}{c}{$A$} &  \multicolumn{1}{c}{$B$} & \multicolumn{1}{c}{$D\expo{3}$} & \multicolumn{1}{c}{$H\expo{7}$} & \multicolumn{1}{c}{$L\expo{11}$} \\
        \hline\noalign{\smallskip}
0   &   0                &  -139.050877(41) &   18.53487308(79) &   1.9089352(64)     &      1.42658(14)      &  -1.4707(11)    \\
1   &   3570.35244(72)   &  -139.32031(62)  &   17.82391990(96) &   1.870447(10)      &      1.37737(33)      &  -1.5700(35)    \\
2   &   6975.09355(69)   &  -139.58940(54)  &   17.1224805(18)  &   1.835828(12)      &      1.31875(33)      &  -1.6991(41)     \\
3   &   10216.1478(12)   &  -139.84424(86)  &   16.427914(26)   &   1.80599(16)       &      1.2447(37)       &  -1.824(32)    \\
4   &   13294.6144(14)   &  -140.0832(17)   &   15.737018(34)   &   1.78267(21)       &      1.1630(37)       &  -2.124(19)        \\
5   &   16210.5997(18)   &  -140.2893(19)   &   15.045630(63)   &   1.76640(58)       &      0.9956(143)      &   [-1.416]         \\
6   &   18962.9921(19)   &  -140.4396(23)   &   14.348665(68)   &   1.76159(67)       &      0.8425(176)      &   [-1.75]            \\
7   &   21549.1777(19)   &  -140.5041(19)   &   13.639345(70)   &   1.77073(73)       &      0.6013(204)      &   [-1.74]            \\
8   &   23964.6390(21)   &  -140.4296(16)   &   12.90888(12)    &   1.8002(18)        &      0.3214(773)      &   [-3.4]            \\
9   &   26202.4452(19)   &  -140.1409(16)   &   12.145349(64)   &   1.85782(47)       &   [-0.111]              &   [-5.8]            \\
10  &   28252.5370(31)   &  -139.5064(17)   &   11.33258(17)    &   1.9571(20)        &   [-0.717]              &   [-12.5]            \\
11  &   30100.8094(29)   &  -138.3238(17)   &   10.45025(33)    &   2.1383(85)        &   [-1.5]                &   ...                \\
12  &   31728.2544(35)   &  -136.3625(19)   &   9.45859(52)     &   2.349(13)         &   [-2.5]                &   ...                \\
13  &   33109.6605(30)   &  -133.0318(23)   &   8.32780(25)     &   2.7906(46)        &   [-4.0]                &   ...                \\
    \hline
    \end{tabular}
    \renewcommand{\tabcolsep}{10pt}
         \begin{tabular}{@{} D{.}{.}{3} @{} D{.}{.}{6} l l@{} D{.}{.}{8} l l@{} D{.}{.}{4} D{.}{.}{8}}
        \mathrm{v}  & \multicolumn{1}{c}{$M\expo{15}$} & \multicolumn{1}{c}{$N\expo{20}$} & \multicolumn{1}{c}{$O\expo{23}$} & \multicolumn{1}{c}{$\gamma\expo{}$} & \multicolumn{1}{c}{$\gamma_D\expo{5}$} &  \multicolumn{1}{c}{$\gamma_H\expo{9}$} & \multicolumn{1}{l}{\ \ $\gamma_L\expo{13}$} & \multicolumn{1}{c}{$q\expo{2}$}  \\
        \hline\noalign{\smallskip}
0   &  0.8842(25)   &   [6.2]     &   [--3.96]  &   -1.19251(14)  &  2.357(13)        &    -3.021(218)  &   3.47(103)   &   -3.86908(17)   \\
1   &  1.258(11)    &   [--8.80]  &   [--4.0]   &   -1.13749(54)  &  2.315(18)        &    -2.467(282)  &   [3.5]      &   -3.69357(18)   \\
2   &  1.721(14)    &   [--32.6]  &   [--4.0]   &   -1.07690(48)  &  2.192(15)        &    -1.523(228)  &   [3.5]      &   -3.51941(29)   \\
3   &  2.007(92)    &   [--66.0]  &   [--4.0]   &   -1.0250(32)   &  2.353(159)       &    -1.96(111)   &   [3.5]      &   -3.3393(21)    \\
4   &   [2.5]       &   [--100]   &   [--4.0]   &   -0.9664(42)   &  2.473(274)       &    -3.14(252)   &   [3.5]      &   -3.1590(33)    \\
5   &   [2.8]       &   [--100]   &   ...       &   -0.9035(46)   &  2.591(266)       &   [--6.4]       &   ...         &  -2.9719(40)     \\
6   &   [3.1]       &   [--100]   &   ...       &   -0.8390(56)   &  3.191(361)       &   [--17.1]      &   ...         &  -2.7800(45)     \\
7   &   [3.4]       &   [--100]   &   ...       &   -0.7593(51)   &  3.240(362)       &   [--17.7]      &   ...         &  -2.5781(49)     \\
8   &   [3.7]       &   [--100]   &   ...       &   -0.6744(63)   &  4.353(576)       &   [--30.0]      &   ...         &  -2.3593(75)     \\
9   &   [4.0]       &   [--100]   &   ...       &   -0.5558(64)   &  5.313(648)       &   [--60]        &   ...         &  -2.1184(82)     \\
10  &   [4.3]       &   [--100]   &   ...       &   -0.4010(85)   &   [6.0]         &   [--100]       &   ...         &     -1.861(29)   \\
11  &   ...           &   ...     &   ...       &   -0.2981(190)  &   [7.0]         &   ...           &   ...         &     -1.564(68)   \\
12  &   ...           &   ...     &   ...       &   0.4404(207)   &   [8.5]         &   ...           &   ...         &     -1.284(31)   \\
13  &   ...           &   ...     &   ...       &   0.8677(115)   &   [10.0]        &   ...           &   ...         &     -0.7239(207) \\
    \hline
    \end{tabular}
    \renewcommand{\tabcolsep}{6pt}
    \begin{tabular}{@{} D{.}{.}{3} @{}  D{.}{.}{9} D{.}{.}{8} D{.}{.}{7} D{.}{.}{4} D{.}{.}{9} D{.}{.}{8} D{.}{.}{7} D{.}{.}{6}}
        \mathrm{v}  & \multicolumn{1}{c}{$q_D\expo{5}$} & \multicolumn{1}{c}{$q_H\expo{9}$} & \multicolumn{1}{c}{$q_L\expo{13}$} & \multicolumn{1}{l}{\ \ $q_M\expo{17}$} & \multicolumn{1}{c}{$p$} &  \multicolumn{1}{c}{$p_D\expo{5}$} & \multicolumn{1}{c}{$p_H\expo{9}$}  & \multicolumn{1}{c}{$p_L\expo{12}$} \\
        \hline\noalign{\smallskip}
0      &   1.4739(14)     &    -2.713(29)      &     4.358(225)  & -3.42(26)            &     0.235355(20)     &    -5.216(21)    &   5.635(380)               &    -1.050(194)      \\
1      &   1.4451(14)     &    -2.591(32)      &     3.961(133)  &  [-2.08]^\mathrm{b}  &     0.224684(30)     &    -5.235(27)    &   7.775(529)               &      [-2.88]^\mathrm{b}    \\
2      &   1.4342(15)     &    -2.786(29)      &     5.180(125)  &  [-2.0]              &     0.213190(81)     &    -5.003(30)    &   4.386(460)               &   [-3.20]^\mathrm{b}    \\
3      &   1.403(13)      &    -2.505(201)     &     4.748(886)  &  [-2.0]              &     0.20276(35)      &    -5.310(233)   &   6.54(191)                &       [-3.0]    \\
4      &   1.400(19)      &    -2.484(137)     &    [4.79]^\mathrm{b}   &  [-2.0]       &     0.19120(43)      &    -5.296(260)   &    [3.80]^\mathrm{b}        &       [-3.0]    \\
5      &   1.388(21)      &    [-2.5]          &    ...          &    ...               &     0.17937(59)      &    -5.701(438)   &      [3.5]                  &    ...     \\
6      &   1.413(26)      &    [-2.75]^\mathrm{b}      &    ...  &    ...               &     0.16642(59)      &    -5.949(485)   &     [2.8]                   &    ...     \\
7      &   1.427(30)      &    [-2.69]^\mathrm{b}      &    ...  &    ...               &     0.15223(63)      &    -6.479(566)   &      [1.8]                  &    ...     \\
8      &   1.464(66)      &    [-3.0]          &    ...          &    ...               &     0.13578(86)      &    -7.111(957)   &      [1.0]                  &    ...     \\
9      &   1.475(79)      &    [-3.3]          &    ...          &    ...               &     0.11694(88)      &    -8.29(116)    &    ...                      &    ...     \\
10     &   1.612(394)     &    [-3.6]         &    ...           &    ...               &     0.09316(231)     &    -8.96(459)    &    ...                      &    ...     \\
11     &      [1.7]^\mathrm{b}    &    ...     &    ...          &    ...               &     0.04981(193)     &    [-12.000]     &    ...                      &    ...     \\
12     &      [1.8]^\mathrm{b}    &    ...     &    ...          &    ...               &     0.03033(168)     &    [-18.000]     &    ...                      &    ...     \\
13     &    [1.9]^\mathrm{b}      &    ...     &    ...          &    ...               &     -0.05720(168)    &    [-27.000]     &    ...                      &    ...     \\
    \hline
        \end{tabular}
    \end{footnotesize}
     \begin{tablenotes}
       \begin{footnotesize}
            \item[a]{Constants in square brackets were fixed; one standard deviation in the last digits is given in parentheses; the constants with their full precision as used by \PGO\ are available in the supplementary material.}
            \item[b]{Fixed here at the value from \citet{2009Bernath-b}, but floated in their fit.}
       \end{footnotesize}
    \end{tablenotes}
  \end{threeparttable}
  }
\end{table*}

\section{Kitt Peak Experimental Spectrum and Herman-Wallis Ratios}\label{sec:Spectrum}
Relative intensities were extracted from an experimental FTS spectrum from \citet{1994Abrams-a}, for later comparison to our calculated values. The spectrum was recorded at the Kitt Peak National Observatory, Arizona, by observing OH emission from the H + O$_3$ reaction, and it has an excellent signal-to-noise ratio. The intensity axis of the observed spectrum was calibrated using a NIST traceable tungsten ribbon ﬁlament lamp (Optronics Model 550C), a spectrum of which had been recorded using the same equipment. The $\Delta \mathrm{v}$=2 region was chosen for analysis as the OH lines are strongest here, it is the cleanest region (mostly due to the presence of water lines in the $\Delta \mathrm{v}$=1 region), and the calibration curve is much flatter in the $\Delta \mathrm{v}$=2 region. Lines were fitted using \textsc{wspectra}~\citep{2001Carleer-a}, the areas were taken as the intensities, and Herman-Wallis (H-W) ratios (described below) were obtained for bands up to 9-7.

OH contains a light H atom, and so exhibits a strong H-W effect~\citep{1955Herman-a} , meaning that the transition intensity heavily depends on $J$. The obvious effect on the spectrum is that the P branches are more intense than the R branches. When comparing calculated intensities to an observed spectrum, this effect can be exploited to validate the line parameters. The intensity ratio of an R branch line to a P branch line with the same upper $J$ level will vary with $J$, and as the same upper (for an emission spectrum) $J$ level is used, the effects on intensity due to energy level population are canceled. This intensity ratio is hereafter referred to as the H-W ratio.

\section{Line Intensity Calculation Method}\label{sec:LineIntensities}
\subsection{Summary}\label{sec:LineIntensities:Summary}
For a full description of the general line intensity calculation procedure, see our other recent work~\citep{2014Brooke-a,2014Brooke-b}, as these calculations were carried out in the same manner.

Briefly, equilibrium constants (Table \ref{tab:EquilibriumConstants}) were calculated by least-squares fitting to the usual power series expansions in v + 1/2 (equations shown in the supplementary material), using the constants $T_\mathrm{v}$ and $B_\mathrm{v}$ from Table \ref{tab:MolecularConstants}. These were employed in the program \RKR\citep{2004LeRoy-Report-a} to generate a potential energy curve, which was then entered into \LEVEL\citep{2007LeRoy-Report-a} along with the DMF. \LEVEL\ generates vibrational wavefunctions and transition matrix elements (MEs), but does not include electron spin. These MEs are therefore in terms of $N$ and not $J$, and we refer to them as Hund's case (b) MEs. They were converted to case (a) MEs which include electron spin using a transformation equation, which was derived in our recent CN work~\citep{2014Brooke-a}, and adjusted here in Section \ref{sec:LineIntensities:a-b}. \PGO\citep{2014Western-Misc-a} sets up standard $N^2$ Hamiltonians using the molecular constants from Table \ref{tab:MolecularConstants}, and adjusts the case (a) transition MEs to include the required averaging over the angles between space and body fixed axis systems. The Hamiltonians are then diagonalized, and the resulting eigenvectors are combined with the transition matrices to give "transformed" transition matrices in terms of the real states. The square of these MEs are the line strengths, $S_{J\primed J}$, which can be used in the following equation to calculate Einstein $A$ values~\citep{2015Bernath-Book-a}:
\begin{eqnarray}\label{eqn:AJJ=SJJ}
A_{J\primed J}
& = &
\frac{16\pi^3 \nu^3 S_{J\primed J}}{3\epsilon_0 hc^3(2J^\prime+1)} \\
& = & 3.136\ 188\ 94 \times 10^{-7} \frac{\tilde{\nu}^3 S_{J\primed J}}{(2J^{\prime}+1)}.
\end{eqnarray}
We report our line intensities in the form of Einstein $A$ values, and also oscillator strengths ($f$-values), which are calculated as follows:
\begin{eqnarray}\label{eqn:fJJ=AJJ}
f_{J\primed J} & = & \frac{m_e \epsilon_0 c^3}{2\pi e^2 \nu^2} \frac{(2J^{\prime}+1)}{ (2J+1)}  A_{J\primed J} \\
 & = & {1.499\ 1938}\frac{1}{\tilde{\nu}^2} \frac{(2J^{\prime}+1)}{(2J+1) } A_{J\primed J},
\end{eqnarray}
where $S_{J\primed J}$ is in debye, and $A_{J\primed J}$ is in seconds.
\begin{table}
\centering
\caption{\label{tab:EquilibriumConstants} Equilibrium molecular constants (in \wavenumbers) for the OH X\DP\ state. These constants with the full precision used in the calculations are available in the online supplementary material. }

  \begin{threeparttable}

     \smallskip
     \begin{footnotesize}
         \begin{tabular}{l @{}  D{.}{.}{15}  l @{}  D{.}{.}{12}}

            \hline\noalign{\smallskip}
            Constant$^\mathrm{a}$ &  \multicolumn{1}{l}{\ \ \ \ \ \ \ \ Value} & Constant &  \multicolumn{1}{l}{\ \ \ \ \ \ \ \ Value} \\
            \hline\noalign{\smallskip}
            $\omega _e$               &  3738.465(19)        &   $B_e$                 &    18.894867(49)    \\
            $\omega _e x_e$           &    84.875(18)        &   $\alpha _{e_1}$       &     0.72343(15)      \\
            $\omega _e y_e$           &     0.5409(77)       &   $\alpha _{e_2}$       &     0.007212(138)    \\
            $\omega _e z_e$           &    -0.02252(170)     &   $\alpha _{e_3}$       &    -0.0006656(469)    \\
            $\omega _e \eta_{e_1}$    &    -0.0009854(1986)  &   $\alpha _{e_4}$       &     0.00005108(598)    \\
            $\omega _e \eta_{e_2}$    &     0.00003087(1155) &   $\alpha _{e_5}$       &    -0.000004828(251)    \\
            $\omega _e \eta_{e_3}$    &    -0.000004539(264)    \\ [1ex]
            \hline
        \end{tabular}
    \end{footnotesize}
     \begin{tablenotes}
       \begin{footnotesize}
       \item[a] Numbers in parentheses indicate one standard deviation to the last significant digits of the constants.
       \end{footnotesize}
    \end{tablenotes}
  \end{threeparttable}
\end{table}
\subsection{Transformation from Hund's case (b) to (a) Matrix Elements}\label{sec:LineIntensities:a-b}

For transition strengths, we require the MEs of the space fixed electric dipole operator, $T^k_p(\mu)$ in spherical tensor notation. This is conventionally expanded~\citep{2003Brown-Book-a} in terms of a molecule fixed dipole moment operator, $T^k_q(\mu)$, and a Wigner $D$  matrix, $D^k_{p,q}(\omega)$, that gives the transformation between the space and molecule fixed axis systems:
\begin{equation}
T^k_p(\mu)=\sum_q D^k_{p,q}(\omega)^*T^k_q(\mu).
\end{equation}
In a Hund's case (a) basis the wavefunctions can be expressed as a product of a rotational part, $|JM\Omega\rangle$,  and a vibronic part, $|\eta \Lambda \mathrm{v}; S \Sigma\rangle$, where $\eta$ represents the remaining electronic and vibrational quantum numbers. Taking matrix elements of the dipole operator, the standard approach yields a product of terms:
\begin{equation}
\sum_q \langle J\primed M\primed \Omega\primed |D^k_{p,q}(\omega)^*|J M \Omega \rangle\langle \eta^\prime \Lambda\primed ;S\primed \Sigma\primed| T^k_q(\bm{\hat{\mu}})|\eta \Lambda ; S \Sigma \rangle,
\end{equation}\label{eqn:ME=3MEs-SphericalTensor-KroneckerSigma}
with the first term containing integrals over the rotational wavefunctions leading to H\"{o}nl-London factors and the second term, the vibronic transition moment being a band strength independent of rotation. To allow for case (a)/(b) mixing the standard methods will give rotational wavefunctions that are linear combinations of the basis functions, giving slightly more complicated expressions.

Given a potential energy curve, $V(r)$, derived as described above, the vibrational wavefunctions, $\Psi(r)$, can be calculated by solving the one dimensional Schr\"{o}dinger equation:
\begin{equation}\label{eqn:1DSch}
\frac{-\hbar^2}{2\mu}\frac{d^2\Psi_{\mathrm{v}}(r)}{dr^2}+V_{\mathrm{v}}(r)\Psi_{\mathrm{v}}(r)=E \Psi_{\mathrm{v}}(r),
\end{equation}
The vibronic transition moment is then evaluated by integrating the DMF, $\mu(r)$ over the vibrational wavefunctions involved:
\begin{equation}\label{eqn:1DSch}
\langle\eta\primed\Lambda\primed|T^k_p(\mu)|\eta\Lambda\rangle=\int^\infty_0\Psi_{\mathrm{v}\primed}(r)\mu(r)\Psi_{\mathrm{v}}(r) dr
\end{equation}
The H-W effect arises because the vibrational wavefunctions change with rotation which can be modelled by adding a centrifugal term:
\begin{equation}\label{eqn:1DSch}
\frac{\hbar^2}{2\mu r^2} \Big(J(J+1)-\Omega^2\Big)=B(r)\Big(J(J+1)-\Omega^2\Big)
\end{equation}
to the potential energy in the Schr\"{o}dinger equation above. A separate vibrational wavefunction, $\Psi_{\mathrm{v},J}(r)$ then arises for each value of $J$, and the formally vibronic transition moment acquires a variation with $J$. To reflect this we add additional labels to the transition moment:
\begin{equation}\label{eqn:1DSch}
\langle\eta\primed\Lambda\primed|T^k_p(\mu, J\primed \Omega\primed J\Omega)|\eta\Lambda\rangle=\int^\infty_0\Psi_{\mathrm{v}\primed, J\primed}(r)\mu(r)\Psi_{\mathrm{v}, J}(r) dr
\end{equation}
This means that the Schr\"{o}dinger equation must be solved separately for each value of $J$.

There are two approaches to taking the spin into account. In the method described by \citet{1992Chackerian-a}, that has been used by \citet{1998Goldman-a} to prepare the line list used in HITRAN 2012, a pair of coupled differential equations are set up, one for $\Omega = 1/2$ and one for $\Omega = 3/2$. This essentially corresponds to a Hund's case (a) basis. The diagonal terms are the one-dimensional Schr\"{o}dinger equation as above, with the addition of a constant spin-orbit coupling term, $A\Lambda\Sigma = \pm A/2$. Note that the spin-orbit coupling constant, $A$, was assumed to be independent of $r$ by \citeauthor{1992Chackerian-a}. This is an approximation, though table \ref{tab:MolecularConstants} indicates a relatively small variation of $A$ with v and thus $r$. The off-diagonal term is $-B(r)\sqrt{J(J+1)-\Omega\Omega\primed}$, analogous to the $-B\hat{J}_+\hat{S}_-$ term normally used to model case (a)/(b) mixing (``spin uncoupling"). Solving the pair of equations gives a vibrational wavefunction for each v, $J$, and $\Omega$, and the vibronic part of the transition moment is then evaluated by integrating the DMF over each pair of wavefunctions on the assumption that it is independent of $\Omega$. While the dipole moment itself should be diagonal in $\Omega$ (with case (a) basis functions), transitions off-diagonal in $\Omega$ will arise by virtue of the case (a)/(b) mixing.

We use a simplified approach here, allowing us to use LeRoy's \LEVEL\ program to solve a single differential equation, and ignoring the electron spin in finding the vibrational wavefunctions. The centrifugal term becomes $B(r)\Big(N(N+1)-\Lambda^2\Big)$ where $N$ is the angular momentum excluding electron spin ($J = N \pm \frac{1}{2}$ here). This will give different results; the most obvious change is that the effective case (a)/(b) mixing term involves an average value of $B$, rather than an operator that depends on $r$. We expect the differences to be small, given that this is anyway the approach used in calculating energy levels, and there are less stringent requirements on the accuracy of intensity calculations. Integrating the dipole moment over wavefunctions derived in this manner gives essentially vibronic transition moments in a case (b) basis, which must be transformed to a case (a) basis for use by \PGO, as it always works in a case (a) basis. The required transformation has been derived and used in our previous papers, but it became clear as part of this work that a small but important correction is required, so we now summarise the revised derivation of the transformation.

The transformation between case (a) and functions, defined in terms of $\Omega = \Lambda + \Sigma$, and case (b) functions, defined in terms of $N$, is~\citep{1976Brown-a}:
\begin{equation}
\begin{split}
& |\eta\Lambda;N\Lambda SJM\rangle = \\
&
\sum_{\Sigma}(-1)^{N-S+\Lambda+\Sigma}\sqrt{2N+1}
\begin{pmatrix}
J & S & N \\
\Lambda+\Sigma & -\Sigma & -\Lambda
\end{pmatrix}
|\eta\Lambda ; S\Sigma ; JM\Omega\rangle.
\end{split}
\end{equation}
This allows any given case (a) matrix element to be expressed in terms of case (b) MEs:
\begin{equation}
\begin{split}
& \langle \eta\primed\Lambda;S\primed\Sigma\primed;J\primed M\primed\Omega\primed| \hat{H} |\eta\Lambda ; S\Sigma ; JM\Omega\rangle  = \\
& \sum_q (-1)^{N\primed-N+S\primed-S+\Omega\primed-\Omega}
\sqrt{(2N\primed+1)(2N+1)}
\begin{pmatrix}
J\primed & S\primed & N\primed \\
\Omega\primed & -\Sigma\primed & -\Lambda\primed
\end{pmatrix}\\
& \times
\begin{pmatrix}
J & S & N \\
\Omega\ & -\Sigma & -\Lambda
\end{pmatrix}
\langle\eta\primed\Lambda\primed ; N\primed \Lambda\primed S \primed J M|\hat{H}|\eta\Lambda;  N \Lambda S J M \rangle.
\end{split}
\end{equation}
The matrix element of the transition dipole moment given in equation (2) above in a Hund's case (a) basis is:
\begin{flalign}
\langle & \eta\primed\Lambda;S\Sigma\primed;J\primed M\primed\Omega\primed| T^k_p(\mu) |\eta\Lambda ; S\Sigma ; JM\Omega\rangle =  \sum_q (-1)^{M\primed-\Omega\primed}&& \\\nonumber
&\times \sqrt{(2J\primed+1)(2J+1)}
\begin{pmatrix}
J\primed & k & J \\
-\Omega\primed & q & \Omega
\end{pmatrix}
\begin{pmatrix}
J\primed & k & J \\
-M\primed & q & M
\end{pmatrix}&& \\\nonumber
& \times
\ \langle\eta\primed\Lambda\primed |T_q^k(J\primed\Omega\primed J\Omega)|\eta\Lambda \rangle. &&
\end{flalign}
and in a Hund's case (b) basis:
\begin{flalign}
\langle & \eta\primed\Lambda;N\primed SJ\primed M\primed| T^k_p(\mu) |\eta\Lambda;N SJM\rangle = (-1)^{J\primed-M\primed}&&\\\nonumber
& \times
\begin{pmatrix}
J\primed & k & J \\
-M\primed & q & M
\end{pmatrix}
(-1)^{N\primed+S+J+k} \sqrt{(2J\primed+1)(2J+1)}&&\\\nonumber
& \times
\begin{Bmatrix}
N^\prime & J^\prime & S \\
J & N & k
\end{Bmatrix}
\sum_q (-1)^{N\primed-\Lambda\primed} \sqrt{(2N\primed+1)(2N+1)}&&\\\nonumber
& \times
\begin{pmatrix}
N\primed & k & N \\
-\Lambda\primed & q & \Lambda
\end{pmatrix}
\langle\eta\primed\Lambda\primed|T_q^k(N\primed N)|\eta\Lambda\rangle.
\end{flalign}\label{eqn:ch-theory:Transformation-case(b)-spin}
In this expression, $\langle\eta\primed\Lambda\primed|T_q^k(N\primed N)|\eta\Lambda\rangle$ is the vibronic transition dipole moment which only depends on $N$ by virtue of the H-W effect, analogous to the $J$ and $\Omega$ dependence in $\langle\eta\primed\Lambda\primed |T_q^k(J\primed\Omega\primed J\Omega)|\eta\Lambda \rangle$. $\langle\eta\primed\Lambda\primed|T_q^k(N\primed N)|\eta\Lambda\rangle$ is the quantity calculated by LeRoy's \LEVEL\ program. Combining the last three equations allows the Hund's case (a) and Hund's case (b) vibronic matrix elements to be related:
\begin{equation}
\begin{split}
\langle & \eta\primed\Lambda\primed|T_{\Omega\primed-\Omega}^k(J\primed\Omega\primed J\Omega)|\eta\Lambda\rangle
 =
 \\
&
(-1) ^{J\primed-\Omega\primed}
\begin{pmatrix}
J\primed & k & J \\
-\Omega\primed & q & \Omega
\end{pmatrix}^{-1}
\sum_{N,N\primed} (-1)^{N\primed-N+\Omega\primed-\Omega+S+J+\Lambda\primed+k}  \\
&\times
(2N\primed+1)(2N+1)
\begin{pmatrix}
J\primed & S & N\primed \\
\Omega\primed & -\Sigma & -\Lambda\primed
\end{pmatrix}
\begin{pmatrix}
J & S & N \\
\Omega & -\Sigma & -\Lambda
\end{pmatrix}
\\
&\times
\begin{Bmatrix}
N^\prime & J^\prime & S \\
J & N & k
\end{Bmatrix}
\begin{pmatrix}
N\primed & k & N \\
-\Lambda\primed & q & \Lambda
\end{pmatrix}
\langle\eta\primed\Lambda\primed|T_{\Lambda\primed-\Lambda}^k(N\primed N)|\eta\Lambda\rangle.
\end{split}
\end{equation}
A more complete derivation is given in the supplementary material. The key difference from the previous version~\citep{2014Brooke-b} is the removal of the condition that $\Omega = \Omega\primed$, and in the OH case this results in matrix elements with $\Delta\Omega = \pm1$ in addition to the $\Delta\Omega = 0$ matrix elements expected in a simple Hund's case (a) basis. That this is a direct consequence of the $N$ (or $J$) dependence of the vibronic transition moment follows from the result (shown in the supplementary information) that if the vibronic transition moment is independent of $N$ the expected result follows:
\begin{equation}\label{}
\langle\eta\primed\Lambda\primed|T_{\Lambda\primed-\Lambda}^k(N\primed N)|\eta\Lambda\rangle=\langle  \eta\primed\Lambda\primed|T_{\Omega\primed-\Omega}^k(J\primed\Omega\primed J\Omega)|\eta\Lambda\rangle,
\end{equation}
and the vibronic transition moment is independent of $J$ and has the selection rule $\Delta\Omega = \Delta\Lambda$ (or $\Delta\Sigma = 0$).

The H-W effect therefore has the effect of introducing small matrix elements off-diagonal in $\Omega$ in the case (a) basis. For example, for the (2,0), R(1.5) transition, the transition matrix changes as shown in Table \ref{tab:TransitionMatrixDeltaSigma}

\begin{table}
\centering
\caption{\label{tab:TransitionMatrixDeltaSigma} Symmetrised \emph{e} transition matrix for the (2,0), R(1.5) example transition  level. Upper - with original transformation equation. Lower - including $\Delta\Sigma\ne0$ MEs. All values are in debye.}

  \begin{threeparttable}

     \smallskip
     \begin{footnotesize}
     \begin{tabular}{ccc}
     \hline\noalign{\smallskip}
       &  $|^2\Pi_{(1.5)}(e/f)\rangle$  & $|^2\Pi_{(0.5)}(e/f)\rangle$   \\
       \noalign{\smallskip} \hline\noalign{\smallskip}
        $|^2\Pi_{(1.5)}(e/f)\rangle$    &       -0.01157  & 0            \\
        $|^2\Pi_{(0.5)}(e/f)\rangle$    &     0   &  -0.01418            \\
        \noalign{\smallskip} \hline\hline\noalign{\smallskip}
    & $|^2\Pi_{(1.5)}(e/f)\rangle$  & $|^2\Pi_{(0.5)}(e/f)\rangle$     \\
        \noalign{\smallskip} \hline\noalign{\smallskip}
     $|^2\Pi_{(1.5)}(e/f)\rangle$  & -0.01157  & -0.0002774            \\
      $|^2\Pi_{(0.5)}(e/f)\rangle$  & -0.0001114  & -0.01418           \\
      \noalign{\smallskip} \hline
        \end{tabular}
    \end{footnotesize}
  \end{threeparttable}
\end{table}

\subsubsection{Comparison to HITRAN Calculations}\label{sec:LineIntensities:a-b:HITRAN}
To show that the results obtained using this adjusted transformation equation are reasonably equivalent to the potentially more accurate HITRAN style calculation involving separate vibrational wavefunctions for each $\Omega$ component, calculations were performed using a DMF equivalent to that used for HITRAN, and also the HITRAN molecular constants. The DMF was constructed using the \citet{1990Nelson-a} DMF at short range (the same range as used for HITRAN), and our ACPF DMF (Section \ref{sec:DMF}) at long range (as the long range part used for HITRAN does not appear to have been published). This means that compared to the actual DMF used for HITRAN, it is identical at short range, but then diverges slightly. Hereafter, it is referred to as the "HITRAN DMF".

\begin{figure}
\centering
   \includegraphics{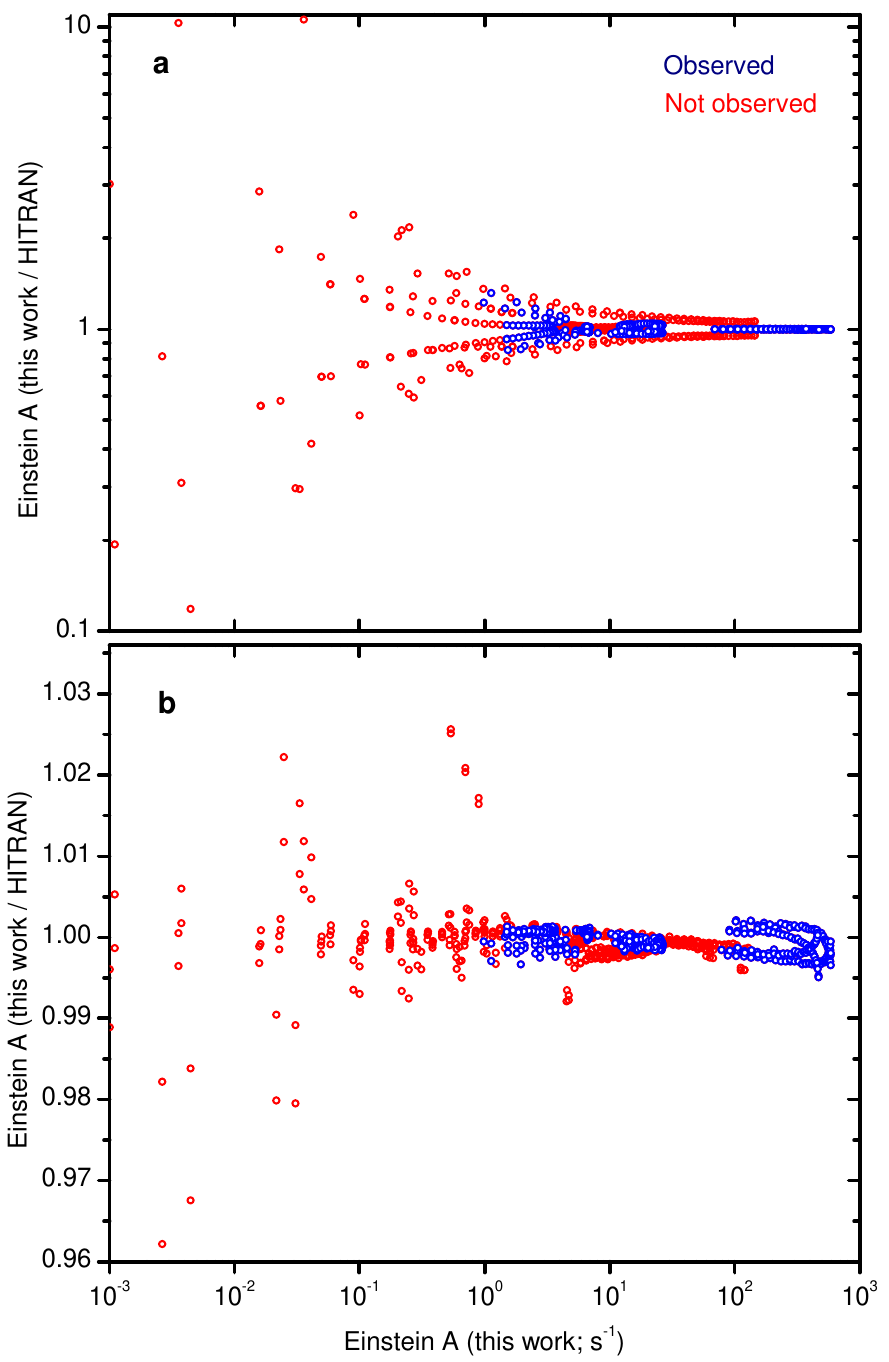}
  \caption{Ratio of Einstein $A$ values between HITRAN and obtained from using the methods described in this work but with the HITRAN 2012 molecular constants and DMF. \textbf{a}: Using the previous version of the transformation equation. \textbf{b}: using the final version of the transformation equation.  Transitions are shown for all bands (including pure rotational) with v$\primed$ up to 2, and where $F\primed=F\Dprimed$. ``Observed" means that the transitions have been identified in experimental spectra, but does not refer to observed intensities. } \label{fig:A-HITRAN}
\end{figure}

Figure \ref{fig:A-HITRAN} plots the ratio of the transition intensities obtained using both the revised and original transformation methods, and the HITRAN DMF and molecular constants. Much better agreement is seen with the use of the revised transformation method. As the potential used in this comparison is not exactly that used for HITRAN, some of the discrepancies are due to to this small difference. Transitions are only shown up to $\mathrm{v}\primed=2$ as the potential and DMF will begin to diverge from those of HITRAN at higher v levels.

\begin{figure}
\centering
   \includegraphics{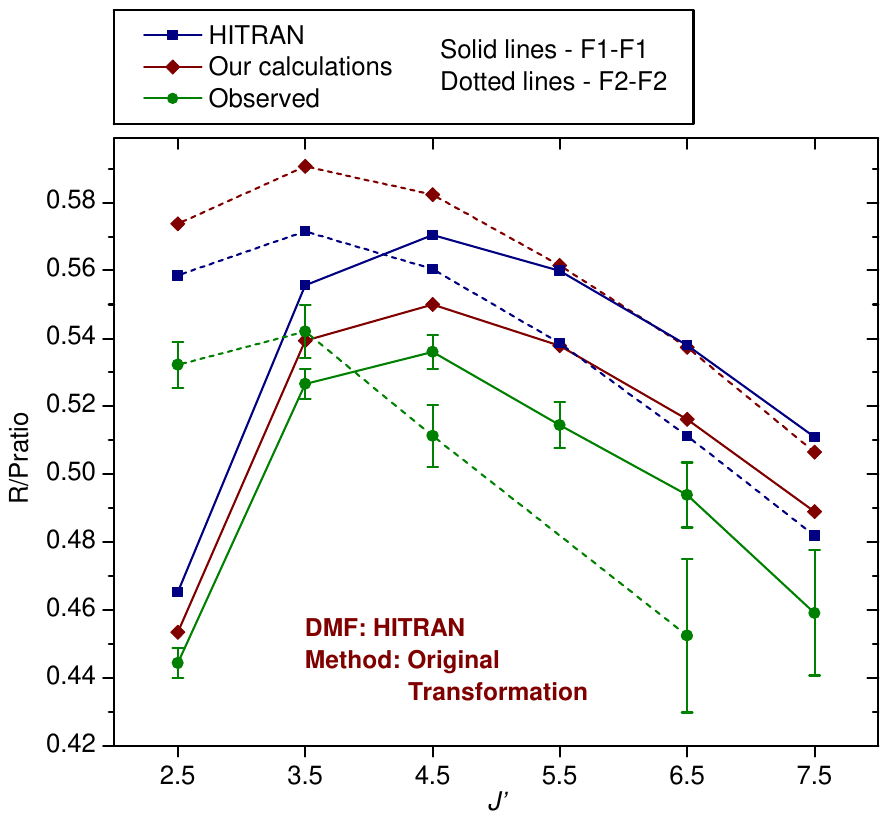}
  \caption{H-W ratios in the OH X\DP\ (2,0) band. The values plotted are equal to the R branch intensity divided by the P branch intensity for transitions that share an upper $J$ and $F$ level. The red lines are calculated using our methods (but the previous version of the transformation equation) and potential, and the HITRAN DMF and molecular constants.} \label{fig:RP-HITRAN}
\end{figure}

The improvement in the transformation method is further demonstrated in Figure \ref{fig:RP-HITRAN}, which compares the H-W ratios obtained with the previous transformation method and HITRAN DMF, with those observed as described in Section \ref{sec:Spectrum}. The main feature in this graph is the splitting of the solid ($F_{11}$ transitions) and dotted ($F_{22}$ transitions) lines of the same color. This splitting is caused by the distribution of intensity between the $F_{11}$ and $F_{22}$ transitions. The splitting pattern of the red lines clearly does not match that of the observed lines, but the splitting of the HITRAN lines matches well, although the HITRAN lines are displaced upwards on the graph from the observed lines (which is caused by inaccuracies in the DMF). With the use of the revised transformation equation, the red lines would match the blue HITRAN lines almost exactly.

\section{Dipole Moment Functions}\label{sec:DMF}
\subsection{Final DMF and Ab Initio Calculations}\label{sec:DMF:AbInitio}
A new DMF was used in this work, which is a combination of two ab initio DMFs. At large internuclear distances, an averaged coupled-pair functional (ACPF)~\citep{1990Werner-a,1988Gdanitz-a,1988Cave-a,1993Szalay-a} DMF was used, and at small distances, a unrestricted open shell coupled-cluster method with single, double, and perturbative triple excitations \big(UCCSD(T)\big)\citep{2010Werner-Misc-a} DMF was used.

The ACPF DMF was calculated using \MOLPRO\ 2012.1~\citep{2012Werner-Misc-a}, using an aug-cc-pV6Z basis set~\citep{1989Dunning-a,1995Woon-a,1996Wilson-a}. This followed a complete active space self-consistent field (CASSCF)~\citep{1985Knowles-a,1985Werner-a} calculation using an active space of four $\sigma$, four $\pi$, and one $\delta$ orbital. The UCCSD(T) DMF was calculated with \MOLPRO\ 2010.1~\citep{2010Werner-Misc-a},  and with the aug-cc-pV6Z basis set\citep{1989Dunning-a,1995Woon-a,1996Wilson-a}. For both DMFs, all core correlation was included, and the dipole moments were calculated by the finite field method.

A comparison of these DMFs and others is shown in Figure \ref{fig:DMFs}. A full description of the mixing of the two DMFs to produce the final version is provided in Section \ref{sec:DMF:Mixing}.

\subsection{Results From Using Only Ab Initio DMFs}\label{sec:DMF:AbInitioResults}
The full calculations were performed using both ab initio DMFs individually, and the results were analysed by comparison to the experimental dipole moments of Peterson \etal\citep{1984Peterson-a}, an experimental lifetime obtained recently by \citet{2005Meerakker-a}, and H-W ratios.

\begin{figure}
  \includegraphics[width=9cm]{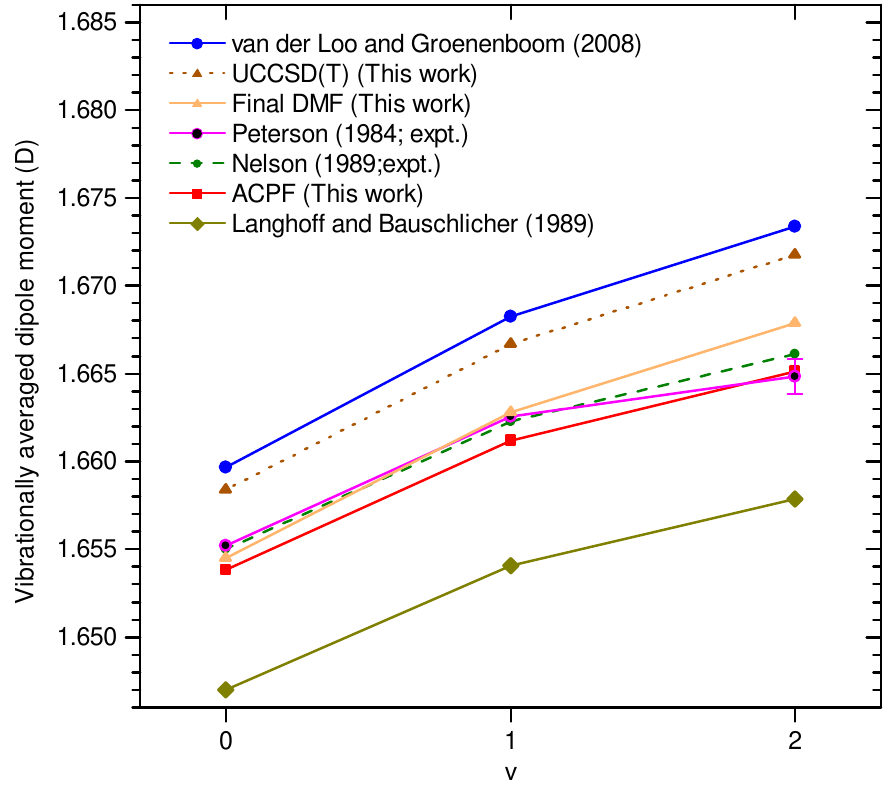}
  \centering
  \caption{Calculated and experimental $\mu_\mathrm{v}$ values of OH (X\DP). The error bars of the experimental values for v=0 and v=1 are not shown as they are slightly smaller than the size of the symbols.} \label{fig:mu_v}
\end{figure}

Better agreement with the Peterson \etal\citep{1984Peterson-a} dipole moments was shown with the ACPF DMF than any of the other DMFs shown, except for that of \citeauthor{1990Nelson-a}, which is expected as the \citeauthor{1990Nelson-a} DMF is partly based on these experimental measurements. This is shown in Figure \ref{fig:mu_v}, where it can also be seen that the UCCSD(T) DMF does not agree as well.

The lifetime of the v=1 level was calculated as the weighted (for $e$ and $f$ parity levels) sum of the Einstein $A$ values for all possible transitions from a single upper $J$ level ($J$=1.5, $F_1$), and taking the reciprocal. Unfortunately, the result of 65.36 ms does not fall within the error bounds of the recent experimental measurement. The UCCSD(T) DMF produces a lifetime in excellent agreement of 59.14 ms. Lifetimes were also calculated using our methods, potential, and line positions, but using the other literature DMFs. All of these are shown in Table \ref{tab:Lifetimes}, which shows that the UCCSD(T) lifetime is the best match to the experimental value.

\begin{table}
\centering
  \begin{threeparttable}
     \caption{\label{tab:Lifetimes} Calculated and experimental lifetimes of the OH X\DP, v=1, $J$=1.5, $F_1$ level, using different DMFs. The first column is the literature source of the DMF. The second is the lifetime reported in that paper, and the third is the lifetime calculated using the specified DMF, but with our methods, potential, and line positions.}
          \smallskip
     \begin{footnotesize}
     \begin{tabular}{l c c}
        \hline\noalign{\smallskip}
        & \multicolumn{2}{c}{Lifetime (ms)} \\
        \cline{2-3}\noalign{\smallskip}
        DMF &  \specialcell{c}{Reported by \\ authors} & \specialcell{c}{Calculated in \\ this work} \\
        \hline\noalign{\smallskip}
        \specialcell{l}{van de Meerakker \\ \etal\ (2007)~\citep{2005Meerakker-a}}  (expt.)       &  59 $\pm$ 2         & ...     \\ [2ex]
        \specialcell{l}{van der Loo and \\ Groenenboom (2008)~\citep{2007vanderLoo-a,2008vanderLoo-a}}        &  56.84              & 56.97   \\ [2ex]
        Langhoff \etal\ (1989)~\citep{1989Langhoff-a}              &  73.3               & 66.7    \\[1ex]
        HITRAN 2012~\citep{2013Rothman-a}   &  56.6 $\pm$ 10-20\% & 56.6    \\ [1ex]
        \specialcell{l}{This work \\ (ACPF DMF)}                          &  ...                & 65.36   \\
        \specialcell{l}{This work \\ (UCCSD(T) DMF)}                          &  ...                & 59.14   \\
        \specialcell{l}{This work \\ \big(Final DMF; UCCSD(T)+ACPF \big)}                          &  ...                & 59.13   \\[1ex]
        \hline
        \end{tabular}
    \end{footnotesize}
  \end{threeparttable}
\end{table}

H-W ratios calculated from the Kitt Peak spectrum as described in Section \ref{sec:Spectrum} were obtained for $\Delta$v=2 bands up to (9,7), and compared to our calculated values and those of HITRAN. Results from the two DMFs are quite similar for the low vibrational bands, and are closer to the observed ratios than HITRAN. This is shown for the 2-0 band as an example in Figure \ref{fig:RP-final}. For the higher vibrational bands, the calculated ratios show worse agreement with the observed ratios for the UCCSD(T) DMF than the ACPF DMF.

\subsection{Mixing of Ab Initio DMFs to Form Final DMF}\label{sec:DMF:Mixing}

The v=1 lifetime is determined mainly by the gradient of the DMF around $R_e$, and as can be seen in Figure \ref{fig:DMFs}, this varies between the various DMFs that have been calculated. The shape of the DMF also determines the magnitude of the H-W effect, and as the UCCSD(T) DMF reproduces these well, its shape is likely to be accurate. The match to the experimental $\mu_\mathrm{v}$ values can be improved by subtracting a constant from the DMF, whilst retaining its shape. The single-reference UCCSD(T) method does not perform well at longer range (in this case above about 2.1 $\AA$). Therefore, it was decided to use the UCCSD(T) at short range, the ACPF at long range, and to mix them in the intermediate region.

The best intermediate region that gave a resulting DMF with smooth first and second derivatives was found to be 1.5 to 2.1 $\AA$. Also to achieve this, a value was added to the ACPF DMF. To ensure that the ACPF still decreased smoothly to zero, this value was smoothly decreased to zero with increasing range.

Specifically, the ACPF DMF was split into three regions, one where the full amount was added ($r < 1.5\ \AA$),  one where the amount added was smoothly decreased to zero ($1.5~\AA~<r~3.0~\AA$), and one where nothing was added ($r > 3.0\ \AA$). The amount added in the smoothing region was equal to
\begin{equation}\label{}
a \frac{1}{2}\Bigg(\cos{\Bigg(\frac{1.5}{\pi}\big(r-1.5\big)}\Bigg)+1\Bigg),
\end{equation}
where $a$ is the maximum amount added.

In the mixing region, the final DMF is a linear combination of the two adjusted ab initio DMFs:
\begin{equation}\label{}
\mathrm{DMF} = c\big(\mathrm{ACPF}\big)+\big(1-c\big)\big(\mathrm{UCCSD(T)}\big),
\end{equation}
where the coefficient $c$ is equal to
\begin{equation}\label{}
\frac{1}{2}\Bigg(\cos{\Bigg(\frac{0.6}{\pi}\big(r-1.5\big)}\Bigg)+1\Bigg).
\end{equation}
A fit of the DMF was performed, in which the final intensities were fitted to the $\Delta$v=2 H-W ratios (from 100 upper $J$ levels) and the \citet{1984Peterson-a} $\mu_\mathrm{v}$ values, and the  parameters floated were the maximum value added to the ACPF DMF (0.04988~D), and the value subtracted from the UCCSD(T) DMF (0.001534~D). The $\mu_\mathrm{v}$ values were weighted such that their weights were proportional to reciprocal of their reported uncertainties, and the sum of their weights was equal to the sum of the weights of all of the H-W ratio data. The H-W ratios were weighted using the reciprocal of their uncertainties (calculated as described in Section \ref{sec:Spectrum}). The v=1 lifetime was not used in the fit, as all of the floated parameters have a negligible effect on it, and it is already reproduced well by the DMF. The mixing of the two DMFs using the final parameters is shown in Figure \ref{fig:MixedDMFs}, and the DMFs are all available in the supplementary material.

\begin{figure}
  \includegraphics[width=9cm]{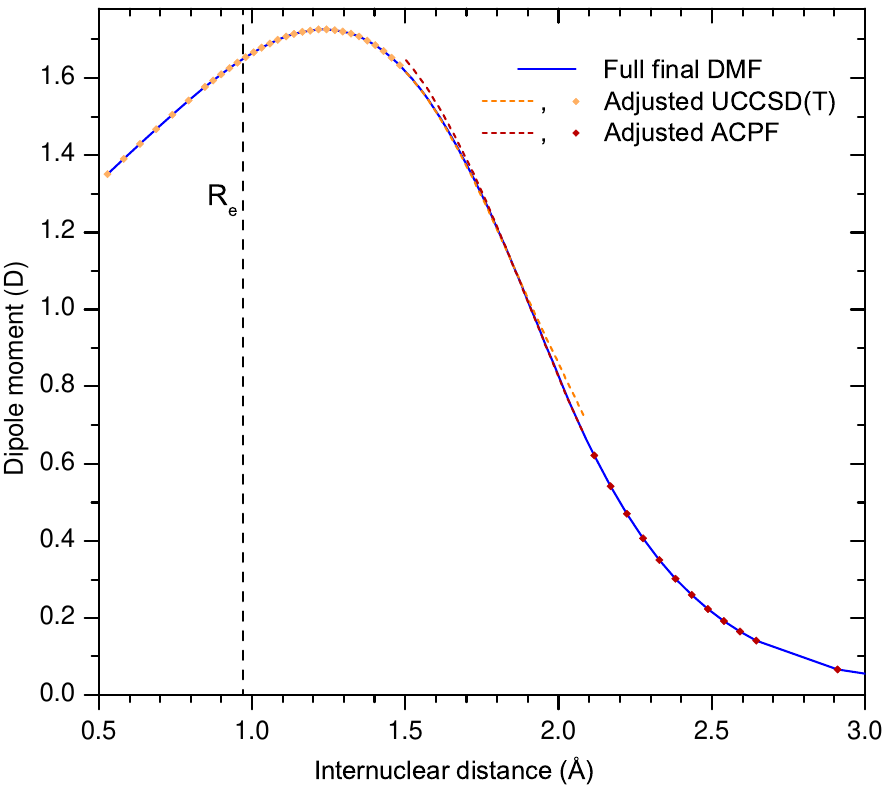}
  \centering
  \caption{Mixing of the two \emph{ab initio} DMFs to form a final DMF.} \label{fig:MixedDMFs}
\end{figure}

\subsection{Comparison to \citet{1990Nelson-a} Fit}\label{sec:DMF:NelsonFit}
\citet{1990Nelson-a} calculated their DMF by fitting an expansion about $R_e$ to three sets of of experimental values: the three \citet{1984Peterson-a} $\mu_\mathrm{v}$ values, about 70 $\Delta$v=1 H-W ratios from their own experimental spectrum, and $\Delta$v=2 to $\Delta$v=1 ratios for nine upper $J$ levels. The uncertainties of their $\Delta$v=1 H-W ratios were larger than for our $\Delta$v=2 H-W ratios, possibly partly due to the $\Delta$v=2 region being much less contaminated by other lines. For the lower vibrational bands, our calculated $\Delta$v=1 H-W ratios match their observed values about as well as theirs do. These differences increase for the higher vibrational bands, but the overall weighted root mean square error using our calculated Einstein $A$ values compared to those of HITRAN is only 1.41 times higher (for $F_{11}$), and 1.22 (for $F_{22}$). For comparison, for the $\Delta$v=2 H-W ratios, these errors are 3.27 and 4.02 times higher using the HITRAN Einstein $A$ values compared to using ours.

\begin{figure}
  \includegraphics[width=9cm]{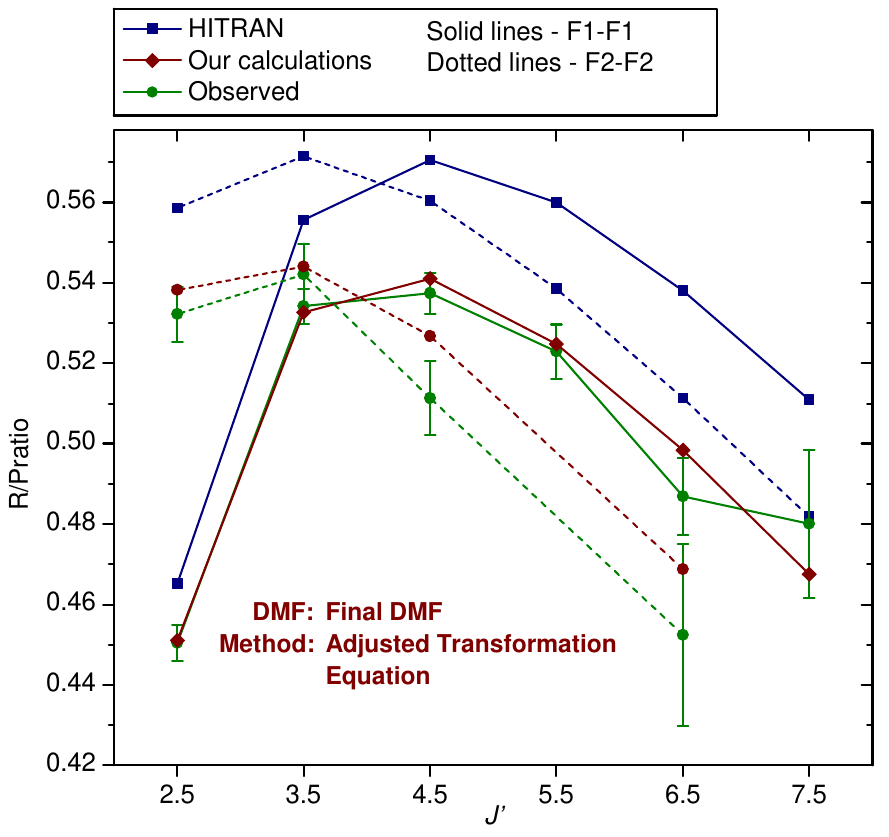}
  \centering
  \caption{Observed and calculated H-W effect in the OH X\DP, (2,0) band, using the final DMF and calculation method. See Figure \ref{fig:RP-HITRAN} for full description.} \label{fig:RP-final}
\end{figure}

A figure equivalent to Figure \ref{fig:RP-HITRAN} was produced using the final data (Figure \ref{fig:RP-final}), which shows a good match to the observed data. Our calculated lifetime (59.13 ms) is also much closer to the experimental measurement of \citet{2005Meerakker-a} than that calculated using the HITRAN DMF (56.6 ms; Table \ref{tab:Lifetimes}).

\section{Final Line List, Lifetimes, and Band Strengths}\label{sec:LineList}
A line list was produced for all possible rovibrational and rotational transitions within the X\DP\ state, for v up to 13, and $J$ up to between 9.5 and 59.5, depending on the band. Transitions are reported between $J$ levels that are a few higher than those observed for each vibrational level.
Vibrational lifetimes were calculated for all of the v levels observed, and are shown in Table \ref{tab:AllLifetimes}.

\begin{table}
\centering
\caption{\label{tab:AllLifetimes} Lifetimes of the first 13 vibrational levels of the OH X\DP\ state.}
  \begin{threeparttable}
          \smallskip
     \begin{footnotesize}
     \begin{tabular}{l @{} D{.}{.}{5}}
        \hline\noalign{\smallskip}
        v & \multicolumn{1}{c}{Lifetime (ms)} \\
        \hline\noalign{\smallskip}
        0     &  59.13   \\
        1     &  31.20   \\
        2     &  21.40   \\
        3     &  16.07   \\
        4     &  12.54   \\
        5     &  10.01   \\
        6     &  8.12    \\
        7     &  6.75    \\
        8     &  5.77    \\
        9     &  5.11    \\
        10    &  4.76    \\
        11    &  4.73    \\
        12    &  5.16    \\  [1ex]
        \hline
        \end{tabular}
    \end{footnotesize}
  \end{threeparttable}
\end{table}

\Avv\ values have been calculated for all reported vibrational bands, and these are available in the online supplementary material. They are calculated by summing over the Einstein $\AJJ$ values for all possible transitions from $J\primed$=1.5, $F\primed$=1, for each band. The \Avv\ values have also been converted into vibrational band oscillator strengths ($\fvv$ -values) using the equation~\citep{1983Larsson-a}

\begin{equation}\label{EQN-fvv=Avv}
\fvv = {1.499 1938\ }\frac{1}{\tilde{\nu}^2}\frac{(2-\delta_{0,\Lambda^\prime})}{(2-\delta_{0,\Lambda^{\prime\prime}) }} A_{\mathrm{v}^\prime \mathrm{v}},
\end{equation}

\noindent where $\Lambda\primed=\Lambda\Dprimed=1$.

\section{Partition Function}\label{sec:Partition}
The partition sums currently provided with the HITRAN database are those from TIPS2011 code described by \citet{2011Laraia-a}. The values are given between 70 and 3000 K. Considering that OH is observed in both very cold and very hot environments, we decided to extend this range to 5-6000 K. As we now have an extended number of calculated energy levels, we calculated the partition function using direct summation over the levels derived in Section \ref{sec:MolecularConstantFit}. Our values agree well with those calculated using TIPS2011 and those provided in the JPL spectroscopic catalogue \big(based on ref. \citep{2013Drouin-a}\big).

\begin{figure}
   \includegraphics[width=9cm]{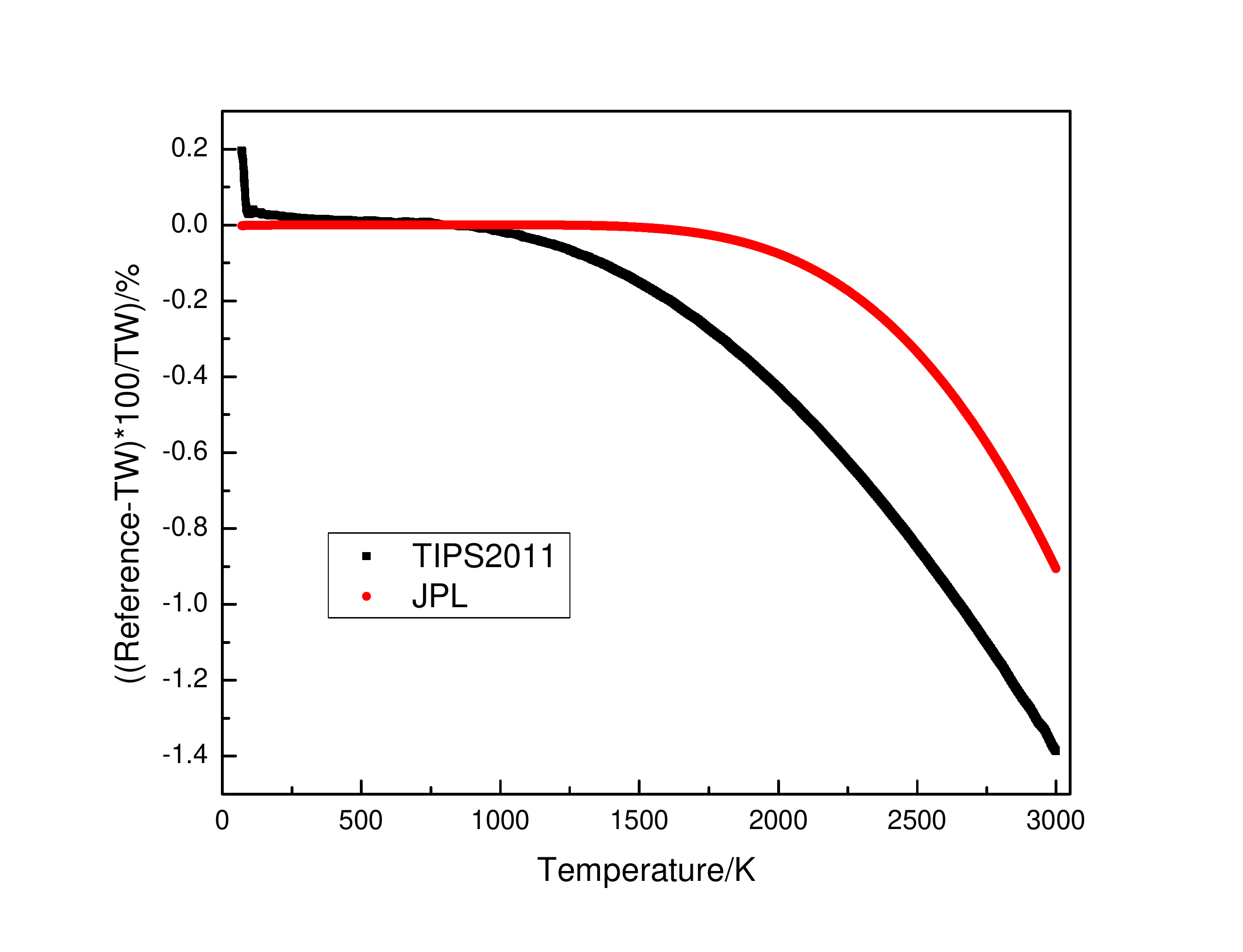}
  \centering
  \caption{Relative differences between the partition sums calculated in this work, and those calculated using TIPS2011 and through direct summation of the energy levels given in the JPL database. TW stands for "this work".} \label{fig:PartitionSum}
\end{figure}

  The hydrogen I=1/2 hyperfine structure is taken into account simply by including a twofold degeneracy for each level, consistent with the approach taken by HITRAN. As the hyperfine splitting is very small, this makes a negligible difference to the calculated values. Figure \ref{fig:PartitionSum} shows the relative differences between partition sums calculated here, and those calculated using TIPS2011 and through direct summation of the energy levels given in the JPL database (which include hyperfine splitting for every level). In general there is good agreement; the discrepancy at low $J$ is an artefact of the interpolation scheme used in TIPS2011, which is based on values tabulated at 25 K intervals. The growing divergence from the JPL values at higher temperatures appears because the JPL catalogue includes only first three vibrational levels. However, in this work we now provide the data between 5-6000 K, which is a significant and important extension to the previously available partition sums. The list of partition sums is given in the supplementary material.

\section{OH and the Oxygen Abundances of the Sun and Stars\label{ohabunds}}

\subsection{General Remarks\label{ohgeneral}}

OH rovibrational bands can be important features for determination of
oxygen abundances in cool stars (those with effective temperatures
T$_{eff}$ $\leq$~5500~K).
They can be attractive alternatives to the visible-region
6300, 6363~\AA\ [O~I] red lines, which are often weak and blended with other
atomic/molecular transitions, and to the 7770~\AA\ high-excitation O~I triplet
lines, whose strengths are subject to significant departures from
local thermodynamic equilibrium (LTE).
Here we report preliminary oxygen abundance analyses from OH lines
for a handful of stars, and compare our results to published
abundances in the literature.

We applied the new OH rovibrational line list to high-resolution spectra
of the Sun and three cool giant stars in the 1.5-1.8~$\mu$m wavelength region
(the astronomical photometric $H$ band).
Many lines of the $\Delta$v=2 system occur in the $H$ band, but a large
fraction of them are blended with transitions of neutral atomic species,
and especially with molecular CN and CO bands.
Therefore for the present exploratory analysis we limited our line list in
this spectral region to 15 relatively clean transitions used in a pioneering
study of the $\Delta$v=2 band system \cite{mel04}.

\begin{figure}
  \includegraphics[width=10cm]{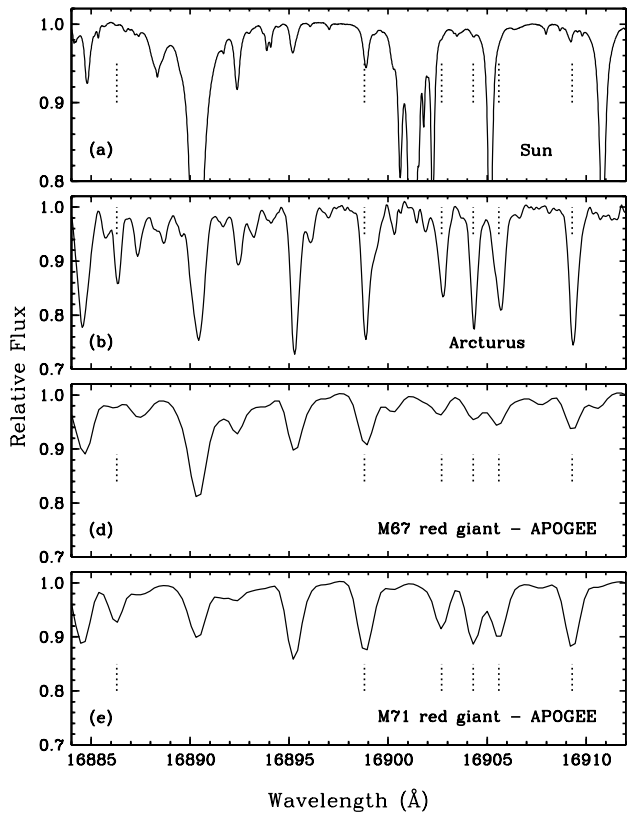}
  \centering
  \caption{Observed spectra of the four program stars in the wavelength
           interval covered by Figure~1 in ref. \cite{mel04}.
           Vertical dotted lines denote wavelengths of the OH lines.
           Telluric lines have been removed from the stellar spectra
           (panels b$-$e) but not from the solar photospheric spectrum
           (panel a).}
\label{fig:Fig-spectra}
\end{figure}

Several large telescopes now have instruments capable of obtaining
high resolution ($R\ \equiv \lambda/\Delta\lambda$ = 20,000$-$80,000), high
signal-to-noise ($S/N$) $H$-band spectra, e.g. VLT CRIRES \citep{kae04},
SDSS APOGEE \citep{wil10}, and newly-commissioned
McDonald IGRINS \citep{par14}.
These facilities will yield a rich set of cool stars with easily detectable OH lines.
The four stars chosen for analysis here represent different temperatures,
gravities, metallicities and Galactic population memberships, and they have
published oxygen abundances.
These stars offer differing challenges for OH analyses.
In Figure~\ref{fig:Fig-spectra} we display a small spectral interval in each
of our program stars.
This wavelength range was chosen to match that in Figure~1 of \citet{mel04}.
The OH lines are strong in Arcturus and the cool giants of M67 and M71,
but are challenging to detect in the other stars.

Lines of the OH $\Delta$v=1 system have been studied as part of a
comprehensive investigation of the solar oxygen abundance \cite{asp04}.
The $\Delta$v=1 lines are much stronger than the $\Delta$v=2 ones, but they
occur in the 3$-$4~$\mu$m spectral domain (the photometric $L$ band).
This bandpass is difficult to access with ground-based high-resolution
spectroscopy because the thermal background is higher and the telluric
H$_2$O absorption is more time-variable in the $L$ band than in the $H$ band.
It is challenging to construct efficient high-resolution spectrographs
for the $L$ band.
Therefore relatively few chemical composition studies have featured
$L$-band data.
However, high-quality spectra are available for the Sun and the mildly
metal-poor giant Arcturus, and we have applied the new line lists to
$\Delta$v=1 lines in these two stars.

In \S\ref{ohstars} we describe the analyses of each star in turn.
Standard stellar spectroscopic notation are employed here:
(a) the ``absolute'' abundances of a given element A in a star is defined as
log $\epsilon$(A) = log $(N_{\mathrm{A}}/N_{\mathrm{H}})$ + 12.0 ;
(b) the relative abundance ratio of elements A and B with respect to their
solar ratio is written as
[A/B] = log $(N_{\mathrm{A}}/N_{\mathrm{B}})_{\star}$ --
log $(N_{\mathrm{A}}/N_{\mathrm{B}})_{\odot}$ ;
and (c), metallicity will be taken to be the [Fe/H] value.

\begin{table*}
\centering
     \caption{\label{tab:absummary} Stellar Model Parameters and Abundance Summary }
  \begin{threeparttable}
     \smallskip
     \begin{scriptsize}
        \begin{tabular}{r c c c D{.}{.}{3} D{.}{.}{4} D{.}{.}{4} D{.}{.}{4}
                        D{.}{.}{4} D{.}{.}{3} D{.}{.}{3} }
        \hline\noalign{\smallskip}
                     Star
                   & T$_{eff}$
                   & log($g$)
                   & v$_{\mathrm{micro}}$
                   & \multicolumn{1}{c}{[Fe/H]}
                   & \multicolumn{1}{c}{log $\epsilon$(C)}
                   & \multicolumn{1}{c}{log $\epsilon$(N)}
                   & \multicolumn{1}{c}{log $\epsilon$(O)}
                   & \multicolumn{1}{c}{log $\epsilon$(O)}
                   & \multicolumn{1}{c}{[O/Fe]}
                   & \multicolumn{1}{c}{[O/Fe]} \\

                   & K
                   &
                   & km s$^{-1}$
                   &
                   & \multicolumn{1}{c}{adopted}
                   & \multicolumn{1}{c}{adopted}
                   & \multicolumn{1}{c}{[O I]}
                   & \multicolumn{1}{c}{OH}
                   & \multicolumn{1}{c}{[O I]}
                   & \multicolumn{1}{c}{OH} \\
        \hline\noalign{\smallskip}
Sun                 & 5780 & 4.44 & 0.85 &  0.00 &   8.43 &   7.83 & 8.69 & 8.72 & +0.00 & +0.03 \\
Arcturus            & 4286 & 1.66 & 2.00 & -0.52 &   8.02 &   7.66 & 8.62 & 8.68 & +0.45 & +0.51 \\
M67$^\mathrm{a}$    & 4623 & 2.44 & 1.62 & +0.06 &   8.38 &   7.94 &  ... & 8.78 &  ... &  +0.03 \\
M71$^\mathrm{b}$    & 4314 & 1.48 & 2.00 & -0.77 & (7.36) & (7.06) &  ... & 8.44 & ... &   +0.52 \\

        \hline\noalign{\smallskip}
    \end{tabular}
    \end{scriptsize}
     \begin{tablenotes}
       \begin{footnotesize}
            \item[a]{star 2M08490674+1129529 in open cluster M67}
            \item[b]{star 2M19533986+1843530 in globular cluster M71}
       \end{footnotesize}
    \end{tablenotes}
  \end{threeparttable}
\end{table*}

\begin{table}
\centering
     \caption{\label{tab:ablines} Parameters and O Abundances from Individual OH Lines }
  \begin{threeparttable}
     \smallskip
     \begin{footnotesize}
        \begin{tabular}{D{.}{.}{6} c c c
                        c c c
                        }
        \hline\noalign{\smallskip}
                     \multicolumn{1}{c}{$\lambda$}
                   & $\chi$
                   & log($gf$)
                   & Sun$^\mathrm{a}$
                   & Arcturus
                   & M67$^\mathrm{^b}$
                   & M71$^\mathrm{^c}$ \\
                     \multicolumn{1}{c}{$\mu$m}
                   & eV
                   &
                   & km s$^{-1}$
                   &
                   &
                   &  \\
        \hline\noalign{\smallskip}
1.52785 & 0.205 & $-$5.435 &  ... & 8.72 & 8.76 & 8.46 \\
1.54091 & 0.255 & $-$5.435 & 8.74 & 8.72 & 8.76 & 8.44 \\
1.55687 & 0.299 & $-$5.337 & 8.74 & 8.67 & 8.78 & 8.45 \\
1.60527 & 0.639 & $-$4.976 & 8.79 & 8.69 & 8.82 & 8.44 \\
1.61921 & 0.688 & $-$4.957 & 8.79 & 8.69 & 8.75 & 8.44 \\
1.63681 & 0.730 & $-$4.858 & 8.74 & 8.65 &  ... & 8.48 \\
1.64560 & 0.609 & $-$5.108 & 8.69 & 8.68 &  ... &  ... \\
1.65345 & 0.781 & $-$4.806 & 8.59 & 8.67 &  ... & 8.42 \\
1.66054 & 0.981 & $-$4.816 & 8.69 & 8.67 & 8.77 & 8.42 \\
1.66559 & 0.681 & $-$5.069 &  ... & 8.70 & 8.83 & 8.47 \\
1.68722 & 0.759 & $-$5.032 & 8.69 &  ... & 8.78 & 8.42 \\
1.68862 & 1.059 & $-$4.720 & 8.74 & 8.62 & 8.75 & 8.41 \\
1.69042 & 0.896 & $-$4.712 &  ... & 8.66 & 8.75 & 8.45 \\
1.69092 & 0.897 & $-$4.712 &  ... & 8.69 & 8.80 & 8.45 \\
1.76189\mathrm{^d} & 1.243 & $-$4.454 &  ... &  ... &  ... &  ...  \\
        &       &          &      &      &      &         \\

         &      \multicolumn{2}{r}{mean\ \ \ \ \ \ \ \ }
                           & 8.72 & 8.68 & 8.78 & 8.44 \\

         &          \multicolumn{2}{r}{+/-\ \ \ \ \ \ \ \     }
                           & 0.02 & 0.01 & 0.01 & 0.01 \\

         &         \multicolumn{2}{r}{sigma\ \ \ \ \ \ \ \ }
                           & 0.06 & 0.03 & 0.03 & 0.02 \\

         &         \multicolumn{2}{r}{\#lines\ \ \ \ \ \ \ \ }
                           &   10 &   13 &   11 &   13 \\
    \hline
        \end{tabular}
    \end{footnotesize}
     \begin{tablenotes}
       \begin{footnotesize}
            \item[a]{all abundances are in units of log $\epsilon$}
            \item[b]{star 2M08490674+1129529 in open cluster M67}
            \item[c]{star 2M19533986+1843530 in globular cluster M71}
            \item[d]{1.76189 $\mu$m from ref. \cite{mel04} was unable to
                     be used in any of our program stars, but is
                     listed here for completeness}
       \end{footnotesize}
    \end{tablenotes}
  \end{threeparttable}
\end{table}

\subsection{Abundances in Individual Stars\label{ohstars}}

\subsubsection{The Solar Photosphere}
We employed the 15 lines given in Table~1 of \citet{mel04} but did not
adopt their equivalent widths (EW) for our solar OH analysis.
These lines are all extremely weak: 0.7~m\AA~$\leq$~EW~$\leq$~1.4~m\AA,  with
$\langle EW \rangle \approx$~1.1~m\AA = 1.1$\times$10$^{-7}$~$\mu$m.
This translates to a mean reduced width of
$\langle RW\rangle \equiv \langle EW\rangle/\lambda$ $\simeq$~$-$7.2.
Such lines are at the limit of reliable $EW$ measurement, and given the
potential for contamination by other solar photospheric and telluric absorptions,
we opted to derive an oxygen abundance for each line by matching synthetic
and observed spectra.
Our analyses followed closely those of previous papers in this series
on MgH \cite{hin13}, C$_2$ \cite{ram14}, and CN \cite{sne14};
here we briefly summarize those methods.

We synthesized a small spectral range, usually 12~$\AA$, surrounding
each OH line, using the LTE stellar spectrum analysis code MOOG
\cite{sne73}\footnote{
Available at http://www.as.utexas.edu/$\sim$chris/moog.html}.
Assembly of the synthesis transition list began with the atomic line
database of \citet{kur11}, adding or updating the line parameters
of atomic species that have had recent lab analyses by the Wisconsin
atomic physics group (\cite{law14} and references therein for Fe-group
elements; \cite{sne09} and references therein for neutron-capture elements).
Molecular transitions of CO \cite{kur11}, CN \cite{sne14} and OH [this
study] were merged with the atomic data.
Since these line lists were also to be applied to the spectra of giant
evolved stars, we include the major CNO isotopologues: $^{12}$CN, $^{13}$CN,
$^{12}$CO, and $^{13}$CO,

For the solar analysis we adopted the photospheric abundances recommended by
\citet{asp09}, in particular log~$\epsilon$(C)~=~8.43,
log~$\epsilon$(N)~=~7.83, log~$\epsilon$(O)~=~8.69,
and $^{12}$C/$^{13}$C~=~89.
To maintain consistency with previous papers in this series, we used the
empirical model photosphere of \citet{hol74}.
Model atmosphere parameters for the Sun and all other stars are given
in Table~\ref{tab:absummary}.
The observed spectrum was that obtained by by L. Delbouille, G. Brault,
J.W. Brault, and L. Testerman at Kitt Peak National Observatory\footnote{
Available at http://bass2000.obspm.fr/solar\_spect.php}.
With the model, line list and observed spectrum inputs we created
synthetic spectra and altered the oxygen abundance until best
synthesis/observation matches were obtained.
Results for each line and the mean abundance statistics are listed in
Table~\ref{tab:ablines}.

Our OH solar photospheric oxygen abundance, log~$\epsilon$(O)~= 8.72, is
only 0.03~dex larger than that recommended by \citet{asp09} from
multiple oxygen-containing species (significantly weighted by the [O~I]
6300~\AA\ transition).
Given the extreme weakness of the OH $\Delta$v=2 solar lines, we regard
the  offset as negligible.
We also examined the photospheric $L$-band spectrum, finding a few very
weak and unblended OH $\Delta$v=1 transitions.
From these we estimate $\langle$log($\epsilon$(O)$\rangle$~$\simeq$~8.75, in
reasonable agreement with the results from the $\Delta$v=2 transitions.

\subsubsection{Thick Disk Red Giant Arcturus}
The very bright mildly metal-poor star Arcturus has been studied many
times at high spectral resolution.
Our recent study of the CN red and violet systems \cite{sne14}
featured Arcturus; that paper gives references to previous analyses.
Its OH lines are strong (panel (b) of Figure~\ref{fig:Fig-spectra}) and
straightforward to analyze.
We used model atmosphere parameters derived by \citet{ram14} to generate
synthetic spectra, and compared them to the Arcturus spectral atlas
\cite{hin00}.
The resulting mean O abundance (Table~\ref{tab:absummary}) of
log~$\epsilon$~=~8.68 translates to a relative overabundance of
[O/Fe]~= $+$0.51.
This value is slightly larger than that derived from the [O~I] lines
by \citet{sne14}, but is in excellent accord with that derived by \citet{ram14}.

\subsubsection{Red Giants in Galactic Open and Globular Clusters}
Members of star clusters are tempting targets for chemical composition
studies since their atmospheric parameters T$_{eff}$ and log~$g$ can often
be reliably estimated simply by broad-band photometry.
APOGEE\footnote{
APOGEE: The ApacheePoint Observatory Galactic Evolution Experiment; see
https://www.sdss3.org/surveys/apogee.php}
\cite{all08} is a dedicated high-resolution spectroscopic survey of more
than 100,000 stars in the $H$ band.
The APOGEE instrument covers most of the $H$ band at a resolving power
of $R$~=~20,000, and currently is producing publicly-available spectra
for both field and cluster red giant targets.
From a list kindly supplied by Gail Zasowsky, we acquired APOGEE spectra
of solar-metallicity old open cluster M67 star 2M08490674+1129529,
and mildly metal-poor globular cluster M71 star 2M19533986+1843530.
The APOGEE database\footnote{
http://data.sdss3.org/bulkIRSpectra/apogeeID}
gives derived atmospheric quantities for these stars that we adopt here
(Table~\ref{tab:absummary}).
These stars do not have previously published O abundances, but the values
derived here are consistent with expectations for their clusters:
(a) [O/Fe]~$\simeq$~0.0 for the the metal-rich M67, and
(b) [O/Fe]~$\simeq$~$+$0.5 for the lower-metallicity M71.
Detailed comparison of O abundances from OH and [O~I] features should
be done in the future, along with derivation of O abundances in all
APOGEE targets in as many clusters as possible.

In summary, the new OH line lists reported in this paper yield O abundances
that are in excellent agreement with those determined from visible
spectral region features. The Einstein $A$ values used in these calculations are around 13\% lower than in HITRAN, which translates to a log(gf) change of -0.06 (from HITRAN to this work). If the calculations were performed with the HITRAN values, the resulting abundances would therefore be approximately 0.06 dex lower. For the Sun, this would result in log $\epsilon$(O) = 8.66, in equally good agreement with the recommended values as our result. The HITRAN value for Arcturus would be log $\epsilon$(O) = 8.66, equal to the value shown in Table \ref{tab:absummary} (calculated by \citet{sne14}), but our value is in better agreement with another recently calculated value (8.76 $\pm0.17$; reported by \citet{2011Ramirez}). Preliminary analyses of OH lines appearing
in new high-resolution IGRINS \citep{par14} spectra of several
evolved stars have also shown better matches to previous abundances for our values than those of HITRAN.

\section{Summary and Conclusion}
New absolute line intensities have been calculated for the OH Meinel system that we believe to be the most accurate available. The absolute line intensities in common use are based on a DMF from 1990~\citep{1990Nelson-a}, and using the newly calculated UCCSD(T)/ACPF mixed DMF, a better match to an experimental lifetime\citep{2005Meerakker-a} and $\Delta$v=2 H-W ratios is observed. Similarly good agreement is seen with the experimental $\mu_\mathrm{v}$ values, but slightly worse agreement with the $\Delta$v=1 H-W ratios observed by \citet{1990Nelson-a}. The new line list includes transitions for v up to 13, and $J$ up to between 9.5 and 59.5, depending on the band, and is available in the online supplementary material.

The ``transformation equation", that converts transition MEs from Hund's case (b) to (a) (and between \LEVEL\ and \PGO) was changed to include $\Delta \Sigma$$\ne$0 MEs. For molecules with a small H-W effect, the effect of this change is very small. It is important when the H-W effect is large such as for OH, but also for NH, for which we recently published a list of intensities~\citep{2014Brooke-b}. The NH list will be updated with the use of this adjusted method in the very near future.

Partition sums have been calculated using the energy levels calculated in this work, and are available in the online supplementary material. They cover the temperature range 5-6000 K, which is an increased range compared to what was previously available in HITRAN (70-3000 K). This will enable the calculation of line intensities in cold environments such as interstellar clouds, and hot environments such as stars.

The absolute intensities for 15 $\Delta$v=2 transitions have been used to calculate  O abundances in the Sun and three other stars, which are  compared to abundances recommended in the literature where available. The OH-based O abundances for the Sun and Arcturus are in agreement with
the values derived from the [O~I] 6300~\AA\ lines, and those for
the two cluster targets are consistent with their expected O abundances.
Additionally, our preliminary analyses of OH lines appearing
in new high-resolution IGRINS \citep{par14} spectra of several
evolved stars, including an extremely metal-deficient star, yield very
reliable O abundances.

The visible spectral region [O~I] lines can be difficult to work with in
many stars, due to contamination by other atomic/molecular stellar features
and by telluric O$_2$ absorption and night-sky [O~I] emission.
Given good high-resolution H-band spectra, the OH lines in some cases
will easily be  able to serve as the primary O abundance indicators,
particularly for target stars that are heavily reddened.

The line list produced in this work will be useful in the fields of atmospheric science, astronomy, and combustion science.

\section{Acknowledgements}

Support for this work was provided by a Research Project Grant from the Leverhulme Trust and a Department of Chemistry (University of York) studentship. Some support was also provided by the NASA Origins of Solar Systems program. MA was supported by the Turkish Scientific and Technological Research Council (T\"{U}B\.{I}TAK, Project 112T929). We are grateful to Gail Zasowski for assistance with the APOGEE database. Thanks also go to Jeremy J. Harrison for helpful discussions about the transformation equation.



\begin{thebibliography}{102}
\providecommand{\natexlab}[1]{#1}
\providecommand{\url}[1]{\texttt{#1}}
\providecommand{\urlprefix}{URL }

\bibitem[{Atkinson and Arey(2003)}]{2003Atkinson-a}
Atkinson R, Arey J.
\newblock Atmospheric degradation of volatile organic compounds.
\newblock Chem Rev 2003;\hspace{0pt}103(12):4605--38.
\newblock \url{http://dx.doi.org/10.1021/cr0206420.}

\bibitem[{{Lelieveld} et~al.(2004){Lelieveld}, {Dentener}, {Peters}, and
  {Krol}}]{2004Lelieveld-a}
{Lelieveld} J, {Dentener} FJ, {Peters} W, {Krol} MC.
\newblock On the role of hydroxyl radicals in the self-cleansing capacity of
  the troposphere 2004;\hspace{0pt}4:2337--44.

\bibitem[{{Prinn} et~al.(1995){Prinn}, {Weiss}, {Miller}, {Huang}, {Alyea},
  {Cunnold}, {Fraser}, {Hartley}, and {Simmonds}}]{1995Prinn-a}
{Prinn} RG, {Weiss} RF, {Miller} BR, {Huang} J, {Alyea} FN, {Cunnold} DM,
  {Fraser} PJ, {Hartley} DE, {Simmonds} PG.
\newblock Atmospheric trends and lifetime of {CH}$_3${CC}l$_3$ and global {OH}
  concentrations.
\newblock Science 1995;\hspace{0pt}269:187--92.
\newblock \url{http://dx.doi.org/10.1126/science.269.5221.187.}

\bibitem[{{Meinel}(1950)}]{1950Meinel-a}
{Meinel} IAB.
\newblock {OH} emission bands in the spectrum of the night sky.
  1950;\hspace{0pt}111:555.
\newblock \url{http://dx.doi.org/10.1086/145296.}

\bibitem[{{Oliva} and {Origlia}(1992)}]{1992Oliva-a}
{Oliva} E, {Origlia} L.
\newblock The {OH} airglow spectrum - a calibration source for infrared
  spectrometers 1992;\hspace{0pt}254:466.

\bibitem[{{Maihara} et~al.(1993){Maihara}, {Iwamuro}, {Yamashita}, {Hall},
  {Cowie}, {Tokunaga}, and {Pickles}}]{1993Maihara-a}
{Maihara} T, {Iwamuro} F, {Yamashita} T, {Hall} DNB, {Cowie} LL, {Tokunaga} AT,
  {Pickles} A.
\newblock Observations of the {OH} airglow emission.
\newblock Astro Soc Pac 1993;\hspace{0pt}105:940--4.
\newblock \url{http://dx.doi.org/10.1086/133259.}

\bibitem[{{Sivjee} and {Hamwey}(1987)}]{1987Sivjee-a}
{Sivjee} GG, {Hamwey} RM.
\newblock Temperature and chemistry of the polar mesopause {OH}
  1987;\hspace{0pt}92:4663--72.
\newblock \url{http://dx.doi.org/10.1029/JA092iA05p04663.}

\bibitem[{{Wilson} et~al.(1972){Wilson}, {Schwartz}, {Neugebauer}, {Harvey},
  and {Becklin}}]{1972Wilson-a}
{Wilson} WJ, {Schwartz} PR, {Neugebauer} G, {Harvey} PM, {Becklin} EE.
\newblock Infrared stars with strong 1665/1667-{MH}z {OH} microwave emission
  1972;\hspace{0pt}177:523.
\newblock \url{http://dx.doi.org/10.1086/151729.}

\bibitem[{{Heiles}(1968)}]{1968Heiles-a}
{Heiles} CE.
\newblock Normal {OH} emission and interstellar dust clouds
  1968;\hspace{0pt}151:919.
\newblock \url{http://dx.doi.org/10.1086/149493.}

\bibitem[{{Piccioni} et~al.(2008){Piccioni}, {Drossart}, {Zasova},
  {Migliorini}, {G{\'e}rard}, {Mills}, {Shakun}, {Garc{\'{\i}}a Mu{\~n}oz},
  {Ignatiev}, {Grassi}, {Cottini}, {Taylor}, {Erard}, and {Virtis-Venus Express
  Technical Team}}]{2008Piccioni-a}
{Piccioni} G, {Drossart} P, {Zasova} L, {Migliorini} A, {G{\'e}rard} JC,
  {Mills} FP, {Shakun} A, {Garc{\'{\i}}a Mu{\~n}oz} A, {Ignatiev} N, {Grassi}
  D, {Cottini} V, {Taylor} FW, {Erard} S, {Virtis-Venus Express Technical
  Team}.
\newblock First detection of hydroxyl in the atmosphere of {V}enus
  2008;\hspace{0pt}483:L29--33.
\newblock \url{http://dx.doi.org/10.1051/0004-6361:200809761.}

\bibitem[{Atreya and Gu(1994)}]{1994Atreya-a}
Atreya SK, Gu ZG.
\newblock Stability of the {M}artian atmosphere: {I}s heterogeneous catalysis
  essential? 1994;\hspace{0pt}99(E6):13133--45.
\newblock \url{http://dx.doi.org/10.1029/94JE01085.}

\bibitem[{{Settersten} et~al.(2003){Settersten}, {Farrow}, and
  {Gray}}]{2003Settersten-a}
{Settersten} TB, {Farrow} RL, {Gray} JA.
\newblock Infrared ultraviolet double-resonance spectroscopy of {OH} in a flame
  2003;\hspace{0pt}369:584--90.
\newblock \url{http://dx.doi.org/10.1016/S0009-2614(03)00022-8.}

\bibitem[{{Maillard} et~al.(1976){Maillard}, {Chauville}, and
  {Mantz}}]{1976Maillard-a}
{Maillard} JP, {Chauville} J, {Mantz} AW.
\newblock High-resolution emission spectrum of {OH} in an oxyacetylene flame
  from 3.7 to 0.9 {$\mu$}m 1976;\hspace{0pt}63:120--41.
\newblock \url{http://dx.doi.org/10.1016/0022-2852(67)90139-7.}

\bibitem[{{Ewart} and {O'Leary}(1986)}]{1986Ewart-a}
{Ewart} P, {O'Leary} SV.
\newblock Detection of {OH} in a flame by degenerate four-wave mixing.
\newblock Opt Lett 1986;\hspace{0pt}11:279--81.
\newblock \url{http://dx.doi.org/10.1364/OL.11.000279.}

\bibitem[{{Abrams} et~al.(1994){Abrams}, {Davis}, {Rao}, {Engleman}, and
  {Brault}}]{1994Abrams-a}
{Abrams} MC, {Davis} SP, {Rao} MLP, {Engleman} R Jr, {Brault} JW.
\newblock High-resolution {F}ourier transform spectroscopy of the {M}einel
  system of {OH} 1994;\hspace{0pt}93:351--95.
\newblock \url{http://dx.doi.org/10.1086/192058.}

\bibitem[{{Grevesse} et~al.(1984){Grevesse}, {Sauval}, and {van
  Dishoeck}}]{1984Grevesse-a}
{Grevesse} N, {Sauval} AJ, {van Dishoeck} EF.
\newblock An analysis of vibration-rotation lines of {OH} in the solar infrared
  spectrum 1984;\hspace{0pt}141:10--6.

\bibitem[{{Mel{\'e}ndez} and {Barbuy}(2002)}]{2002Melendez-a}
{Mel{\'e}ndez} J, {Barbuy} B.
\newblock Keck {NIRSPEC} infrared {OH} lines: {O}xygen abundances in metal-poor
  stars down to [{F}e/{H}] = -2.9 2002;\hspace{0pt}575:474--83.
\newblock \url{http://dx.doi.org/10.1086/341142.}

\bibitem[{{Smith} et~al.(2013){Smith}, {Cunha}, {Shetrone}, {Meszaros},
  {Allende Prieto}, {Bizyaev}, {Garc{\`i}a P{\`e}rez}, {Majewski}, {Schiavon},
  {Holtzman}, and {Johnson}}]{2013Smith-a}
{Smith} VV, {Cunha} K, {Shetrone} MD, {Meszaros} S, {Allende Prieto} C,
  {Bizyaev} D, {Garc{\`i}a P{\`e}rez} A, {Majewski} SR, {Schiavon} R,
  {Holtzman} J, {Johnson} JA.
\newblock Chemical abundances in field red giants from high-resolution {H}-band
  spectra using the {APOGEE} spectral linelist 2013;\hspace{0pt}765:16.
\newblock \url{http://dx.doi.org/10.1088/0004-637X/765/1/16.}

\bibitem[{{Stevens} et~al.(1974){Stevens}, {Das}, {Wahl}, {Krauss}, and
  {Neumann}}]{1974Stevens-a}
{Stevens} WJ, {Das} G, {Wahl} AC, {Krauss} M, {Neumann} D.
\newblock Study of the ground state potential curve and dipole moment of {OH}
  by the method of optimized valence configurations
  1974;\hspace{0pt}61:3686--99.
\newblock \url{http://dx.doi.org/10.1063/1.1682554.}

\bibitem[{{Werner} et~al.(1983){Werner}, {Rosmus}, and
  {Reinsch}}]{1983Werner-a}
{Werner} HJ, {Rosmus} P, {Reinsch} EA.
\newblock Molecular properties from {MCSCF-SCEP} wave functions. {I}.
  {A}ccurate dipole moment functions of {OH}, {OH}$^{ - }$, and {OH}$^{ + }$
  1983;\hspace{0pt}79:905--16.
\newblock \url{http://dx.doi.org/10.1063/1.445867.}

\bibitem[{{Langhoff} et~al.(1986){Langhoff}, {Werner}, and
  {Rosmus}}]{1986Langhoff-a}
{Langhoff} SR, {Werner} HJ, {Rosmus} P.
\newblock Theoretical transition probabilities for the {OH} {M}einel system
  1986;\hspace{0pt}118:507--29.
\newblock \url{http://dx.doi.org/10.1016/0022-2852(86)90186-4.}

\bibitem[{Langhoff et~al.(1989)Langhoff, Bauschlicher, and
  Taylor}]{1989Langhoff-a}
Langhoff SR, Bauschlicher CW, Taylor PR.
\newblock Theoretical study of the dipole moment function of {OH}
  ({X}$^2${$\Pi$}).
\newblock J Chem Phys 1989;\hspace{0pt}91(10):5953--59.
\newblock \url{http://dx.doi.org/10.1063/1.457413.}

\bibitem[{{van der Loo} and {Groenenboom}(2007)}]{2007vanderLoo-a}
{van der Loo} MPJ, {Groenenboom} GC.
\newblock Theoretical transition probabilities for the {OH} {M}einel system
  2007;\hspace{0pt}126(11):114314.
\newblock \url{http://dx.doi.org/10.1063/1.2646859.}

\bibitem[{van~der Loo and Groenenboom(2008)}]{2008vanderLoo-a}
van~der Loo MPJ, Groenenboom GC.
\newblock Erratum: {T}heoretical transition probabilities for the {OH} {M}einel
  system ({J}. {C}hem. {P}hys. 126, 114314 (2007)).
\newblock J Chem Phys 2008;\hspace{0pt}128(15):159902.
\newblock \url{http://dx.doi.org/10.1063/1.2899016.}

\bibitem[{Peterson et~al.(1984)Peterson, Fraser, and
  Klemperer}]{1984Peterson-a}
Peterson KI, Fraser GT, Klemperer W.
\newblock Electric dipole moment of {X$^2\Pi$} {OH} and {OD} in several
  vibrational states.
\newblock Can J Phys 1984;\hspace{0pt}62(12):1502--07.
\newblock \url{http://dx.doi.org/10.1139/p84-196.}

\bibitem[{Nelson et~al.(1990)Nelson, Schiffman, Nesbitt, Orlando, and
  Burkholder}]{1990Nelson-a}
Nelson DD, Schiffman A, Nesbitt DJ, Orlando JJ, Burkholder JB.
\newblock H+{O$_3$} {F}ourier-transform infrared emission and laser absorption
  studies of {OH} {(X$^2\Pi$)} radical: {A}n experimental dipole moment
  function and state-to-state {E}instein {$A$} coefficients.
\newblock J Chem Phys 1990;\hspace{0pt}93(10):7003--19.
\newblock \url{http://dx.doi.org/10.1063/1.459476.}

\bibitem[{{Nelson} et~al.(1989){Nelson}, {Schiffman}, {Nesbitt}, and
  {Yaron}}]{1989Nelson-a}
{Nelson} DD Jr, {Schiffman} A, {Nesbitt} DJ, {Yaron} DJ.
\newblock Absolute infrared transition moments for open shell diatomics from
  {$J$} dependence of transition intensities - application to {OH}
  1989;\hspace{0pt}90:5443--54.
\newblock \url{http://dx.doi.org/10.1063/1.456450.}

\bibitem[{{Turnbull} and {Lowe}(1988)}]{1988Turnbull}
{Turnbull} DN, {Lowe} RP.
\newblock An empirical determination of the dipole moment function of {OH}
  {(X\DP)} 1988;\hspace{0pt}89:2763--7.
\newblock \url{http://dx.doi.org/10.1063/1.455028.}

\bibitem[{{Mies}(1974)}]{1974Mies-a}
{Mies} FH.
\newblock Calculated vibrational transition probabilities of
  {OH}({X}$^{2}${$\Pi$}) 1974;\hspace{0pt}53:150--88.
\newblock \url{http://dx.doi.org/10.1016/0022-2852(74)90125-8.}

\bibitem[{{Goldman} et~al.(1998){Goldman}, {Schoenfeld}, {Goorvitch},
  {Chackerian}, {Dothe}, {M{\'e}len}, {Abrams}, and {Selby}}]{1998Goldman-a}
{Goldman} A, {Schoenfeld} WG, {Goorvitch} D, {Chackerian} C Jr, {Dothe} H,
  {M{\'e}len} F, {Abrams} MC, {Selby} JEA.
\newblock Updated line parameters for {OH} {X}$^{2}${II-X}$^{2}${II}
  ($\upsilon\primed\upsilon$) transitions. 1998;\hspace{0pt}59:453--69.
\newblock \url{http://dx.doi.org/10.1016/S0022-4073(97)00112-X.}

\bibitem[{{Rothman} et~al.(2013){Rothman}, {Gordon}, {Babikov}, {Barbe}, {Chris
  Benner}, {Bernath}, {Birk}, {Bizzocchi}, {Boudon}, {Brown}, {Campargue},
  {Chance}, {Cohen}, {Coudert}, {Devi}, {Drouin}, {Fayt}, {Flaud}, {Gamache},
  {Harrison}, {Hartmann}, {Hill}, {Hodges}, {Jacquemart}, {Jolly}, {Lamouroux},
  {Le Roy}, {Li}, {Long}, {Lyulin}, {Mackie}, {Massie}, {Mikhailenko},
  {M{\"u}ller}, {Naumenko}, {Nikitin}, {Orphal}, {Perevalov}, {Perrin},
  {Polovtseva}, {Richard}, {Smith}, {Starikova}, {Sung}, {Tashkun}, {Tennyson},
  {Toon}, {Tyuterev}, and {Wagner}}]{2013Rothman-a}
{Rothman} LS, {Gordon} IE, {Babikov} Y, {Barbe} A, {Chris Benner} D, {Bernath}
  PF, {Birk} M, {Bizzocchi} L, {Boudon} V, {Brown} LR, {Campargue} A, {Chance}
  K, {Cohen} EA, {Coudert} LH, {Devi} VM, {Drouin} BJ, {Fayt} A, {Flaud} JM,
  {Gamache} RR, {Harrison} JJ, {Hartmann} JM, {Hill} C, {Hodges} JT,
  {Jacquemart} D, {Jolly} A, {Lamouroux} J, {Le Roy} RJ, {Li} G, {Long} DA,
  {Lyulin} OM, {Mackie} CJ, {Massie} ST, {Mikhailenko} S, {M{\"u}ller} HSP,
  {Naumenko} OV, {Nikitin} AV, {Orphal} J, {Perevalov} V, {Perrin} A,
  {Polovtseva} ER, {Richard} C, {Smith} MAH, {Starikova} E, {Sung} K, {Tashkun}
  S, {Tennyson} J, {Toon} GC, {Tyuterev} VG, {Wagner} G.
\newblock The {HITRAN} 2012 molecular spectroscopic database
  2013;\hspace{0pt}130:4--50.
\newblock \url{http://dx.doi.org/10.1016/j.jqsrt.2013.07.002.}

\bibitem[{{Rothman} et~al.(2010){Rothman}, {Gordon}, {Barber}, {Dothe},
  {Gamache}, {Goldman}, {Perevalov}, {Tashkun}, and {Tennyson}}]{2010Rothman-a}
{Rothman} LS, {Gordon} IE, {Barber} RJ, {Dothe} H, {Gamache} RR, {Goldman} A,
  {Perevalov} VI, {Tashkun} SA, {Tennyson} J.
\newblock {HITEMP}, the high-temperature molecular spectroscopic database
  2010;\hspace{0pt}111:2139--50.
\newblock \url{http://dx.doi.org/10.1016/j.jqsrt.2010.05.001.}

\bibitem[{{Bernath} and {Colin}(2009)}]{2009Bernath-b}
{Bernath} PF, {Colin} R.
\newblock Revised molecular constants and term values for the {X}$^{2}${$\Pi$}
  and {B}$^{2}${$\Sigma$} $^{+}$ states of {OH} 2009;\hspace{0pt}257:20--3.
\newblock \url{http://dx.doi.org/10.1016/j.jms.2009.06.003.}

\bibitem[{{Melen} et~al.(1995){Melen}, {Sauval}, {Grevesse}, {Farmer},
  {Servais}, {Delbouille}, and {Roland}}]{1995Melen-a}
{Melen} F, {Sauval} AJ, {Grevesse} N, {Farmer} CB, {Servais} C, {Delbouille} L,
  {Roland} G.
\newblock A new analysis of the {OH} radical spectrum from solar infrared
  observations. 1995;\hspace{0pt}174:490--509.
\newblock \url{http://dx.doi.org/10.1006/jmsp.1995.0018.}

\bibitem[{{Nizkorodov} et~al.(2001){Nizkorodov}, {Harper}, and
  {Nesbitt}}]{2001Nizkorodov-a}
{Nizkorodov} SA, {Harper} WW, {Nesbitt} DJ.
\newblock Fast vibrational relaxation of {OH} (v=9) by ammonia and ozone
  2001;\hspace{0pt}341:107--14.
\newblock \url{http://dx.doi.org/10.1016/S0009-2614(01)00371-2.}

\bibitem[{{Sappey} and {Copeland}(1990)}]{1990Sappey-a}
{Sappey} AD, {Copeland} RA.
\newblock Laser double-resonance study of {OH} {(X$^{2}\Pi_i$}, v~=~12)
  1990;\hspace{0pt}143:160--8.
\newblock \url{http://dx.doi.org/10.1016/0022-2852(90)90267-T.}

\bibitem[{{Copeland} et~al.(1993){Copeland}, {Chalamala}, and
  {Coxon}}]{1993Copeland-a}
{Copeland} RA, {Chalamala} BR, {Coxon} JA.
\newblock Laser-induced fluorescence of the
  {B}$^{2}${$\Sigma$}$^{+}$-{X}$^{2}${$\Pi$} system of {OH}: {M}olecular
  {C}onstants for {B}$^{2}${$\Sigma$}$^{+}$ (v=0,1) and {X}$^{2}${$\Pi$}
  (v=7-9, 11-13) 1993;\hspace{0pt}161:243--52.
\newblock \url{http://dx.doi.org/10.1006/jmsp.1993.1229.}

\bibitem[{{Abrams} et~al.(1996){Abrams}, {Goldman}, {Gunson}, {Rinsland}, and
  {Zander}}]{1996Abrams-a}
{Abrams} MC, {Goldman} A, {Gunson} MR, {Rinsland} CP, {Zander} R.
\newblock Observations of the infrared solar spectrum from space by the {ATMOS}
  experiment.
\newblock App Optics 1996;\hspace{0pt}35:2747--51.
\newblock \url{http://dx.doi.org/10.1364/AO.35.002747.}

\bibitem[{{Farmer} and {Norton}(1989)}]{1989Farmer-Book-a}
{Farmer} CB, {Norton} RH.
\newblock A high-resolution atlas of the infrared spectrum of the {S}un and the
  {E}arth atmosphere from space. A compilation of {ATMOS} spectra of the region
  from 650 to 4800 cm$^{-1}$ (2.3 to 16 {$\mu$}m). {V}ol. {I}. {T}he {S}un.
\newblock National Aeronautics and Space Administration, Hampton, VA,
  USA.~Langley Research Center, 1989.

\bibitem[{{Hardwick} and {Whipple}(1991)}]{1991Hardwick-a}
{Hardwick} JL, {Whipple} GC.
\newblock Far infrared emission spectrum of the {OH} radical
  1991;\hspace{0pt}147:267--73.
\newblock \url{http://dx.doi.org/10.1016/0022-2852(91)90185-D.}

\bibitem[{{Varberg} and {Evenson}(1993)}]{1993Varberg-a}
{Varberg} TD, {Evenson} KM.
\newblock The rotational spectrum of {OH} in the v=0-3 levels of its ground
  state 1993;\hspace{0pt}157:55--67.
\newblock \url{http://dx.doi.org/10.1006/jmsp.1993.1005.}

\bibitem[{{Brown} et~al.(1986){Brown}, {Zink}, {Jennings}, {Evenson}, and
  {Hinz}}]{1986Brown-a}
{Brown} JM, {Zink} LR, {Jennings} DA, {Evenson} KM, {Hinz} A.
\newblock Laboratory measurement of the rotational spectrum of the {OH} radical
  with tunable far-infrared research 1986;\hspace{0pt}307:410--3.
\newblock \url{http://dx.doi.org/10.1086/164427.}

\bibitem[{{Davies} et~al.(1979){Davies}, {Hack}, {Preuss}, and
  {Temps}}]{1979Davies-a}
{Davies} PB, {Hack} W, {Preuss} AW, {Temps} F.
\newblock Far infrared laser magnetic resonance spectra of vibrationally
  excited {OH} 1979;\hspace{0pt}64:94--7.
\newblock \url{http://dx.doi.org/10.1016/0009-2614(79)87283-8.}

\bibitem[{{Radford}(1968)}]{1968Radford-a}
{Radford} HE.
\newblock Scanning microwave echo box spectrometer
  1968;\hspace{0pt}39:1687--91.
\newblock \url{http://dx.doi.org/10.1063/1.1683203.}

\bibitem[{{Ball} et~al.(1970){Ball}, {Dickinson}, {Gottlieb}, and
  {Radford}}]{1970Ball-a}
{Ball} JA, {Dickinson} DF, {Gottlieb} CA, {Radford} HE.
\newblock The 3.8-cm spectrum of {OH}: {L}aboratory measurement and low-noise
  search in {W3} {(OH)} 1970;\hspace{0pt}75:762.
\newblock \url{http://dx.doi.org/10.1086/111022.}

\bibitem[{{Meerts} and {Dymanus}(1975)}]{1975Meerts-a}
{Meerts} WL, {Dymanus} A.
\newblock A molecular beam electric resonance study of the hyperfine
  {$\Lambda$} doubling spectrum of {OH}, {OD}, {SH}, and {SD}
  1975;\hspace{0pt}53:2123.
\newblock \url{http://dx.doi.org/10.1139/p75-261.}

\bibitem[{{Coxon} et~al.(1979){Coxon}, {Sastry}, {Austin}, and
  {Levy}}]{1979Coxon-a}
{Coxon} JA, {Sastry} KVLN, {Austin} JA, {Levy} DH.
\newblock The microwave spectrum of the {OH} {X}$^{2}${$\Pi$} radical in the
  ground and vibrationally-excited ({$\nu$}{$\le$} 6) levels.
  1979;\hspace{0pt}57:619--34.
\newblock \url{http://dx.doi.org/10.1139/p79-089.}

\bibitem[{{Destombes} et~al.(1974){Destombes}, {Marli{\`e}re}, {Rohart}, and
  {Burie}}]{1974Destombes-a}
{Destombes} JL, {Marli{\`e}re} C, {Rohart} F, {Burie} J.
\newblock Nouvelle analyse du spectre hertzien du radical hydroxyl.
\newblock C R Acad Sci B 1974;\hspace{0pt}278:275--8.

\bibitem[{{Destombes} et~al.(1975){Destombes}, {Journel}, {Marli{\`e}re}, and
  {Rohart}}]{1975Destombes-a}
{Destombes} JL, {Journel} G, {Marli{\`e}re} C, {Rohart} F.
\newblock Microwave spectrum of hydroxyl radical in {$\Pi$}-2(3/2) and
  {$\Pi$}-2(1).
\newblock C R Acad Sci B 1975;\hspace{0pt}280:809--11.

\bibitem[{{Kolbe} et~al.(1981){Kolbe}, {Zollner}, and {Leskovar}}]{1981Kolbe-a}
{Kolbe} WF, {Zollner} WD, {Leskovar} B.
\newblock Microwave spectrometer for the detection of transient gaseous species
  1981;\hspace{0pt}52:523--32.
\newblock \url{http://dx.doi.org/10.1063/1.1136633.}

\bibitem[{{ter Meulen} and {Dymanus}(1972)}]{1972terMeulen-a}
{ter Meulen} JJ, {Dymanus} A.
\newblock Beam-maser measurements of the ground-state transition frequencies of
  {OH}. 1972;\hspace{0pt}172:L21.
\newblock \url{http://dx.doi.org/10.1086/180882.}

\bibitem[{{ter Meulen} et~al.(1976){ter Meulen}, {Meerts}, {van Mierlo}, and
  {Dymanus}}]{1976terMeulen-b}
{ter Meulen} JJ, {Meerts} WL, {van Mierlo} GW, {Dymanus} A.
\newblock Observations of population inversion between the {$\Lambda$}-doublet
  states of {OH} 1976;\hspace{0pt}36:1031--4.
\newblock \url{http://dx.doi.org/10.1103/PhysRevLett.36.1031.}

\bibitem[{{Destombes} and {Marliere}(1975)}]{1975Destombes-b}
{Destombes} JL, {Marliere} C.
\newblock Measurement of hyperfine splitting in the {OH} radical by a
  radio-frequency microwave double resonance method 1975;\hspace{0pt}34:532--6.
\newblock \url{http://dx.doi.org/10.1016/0009-2614(75)85556-4.}

\bibitem[{{ter Meulen}(1976)}]{1976terMeulen-a}
{ter Meulen} JJ.
\newblock Ph.D. thesis, Katholieke Universiteit, Nijmegen, The Netherlands,
  1976.

\bibitem[{{Meerts} et~al.(1979){Meerts}, {Bekooy}, and
  {Dymanus}}]{1979Meerts-a}
{Meerts} WL, {Bekooy} JP, {Dymanus} A.
\newblock Vibrational effects in the hydroxyl radical by molecular beam
  electric resonance spectroscopy 1979;\hspace{0pt}37:425--39.
\newblock \url{http://dx.doi.org/10.1080/00268977900100361.}

\bibitem[{{Destombes} et~al.(1979){Destombes}, {Lemoine}, and
  {Marliere-Demuynck}}]{1979Destombes-a}
{Destombes} JL, {Lemoine} B, {Marliere-Demuynck} C.
\newblock Microwave spectrum of the hydroxyl radical in the v=1,
  $^{2}${$\Pi$}$_{3/2}$ state. 1979;\hspace{0pt}60:493--5.
\newblock \url{http://dx.doi.org/10.1016/0009-2614(79)80619-3.}

\bibitem[{{Clough} et~al.(1971){Clough}, {Curran}, and {Thrush}}]{1971Clough-a}
{Clough} PN, {Curran} AH, {Thrush} BA.
\newblock The {E.P.R.} spectrum of vibrationally excited hydroxyl radicals
  1971;\hspace{0pt}323:541--54.
\newblock \url{http://dx.doi.org/10.1098/rspa.1971.0122.}

\bibitem[{{Lee} et~al.(1971){Lee}, {Tam}, {Larouche}, and
  {Woonton}}]{1971Lee-a}
{Lee} KP, {Tam} WG, {Larouche} R, {Woonton} GA.
\newblock Electron resonance of vibrationally excited {OH} radicals.
  1971;\hspace{0pt}49:2207--10.
\newblock \url{http://dx.doi.org/10.1139/p71-267.}

\bibitem[{Martin-Drumel et~al.(2011)Martin-Drumel, Pirali, Balcon, Bréchignac,
  Roy, and Vervloet}]{2011Martin-Drumel-a}
Martin-Drumel MA, Pirali O, Balcon D, Bréchignac P, Roy P, Vervloet M.
\newblock High resolution far-infrared {F}ourier transform spectroscopy of
  radicals at the {AILES} beamline of {SOLEIL} synchrotron facility
  2011;\hspace{0pt}82(11):113106.
\newblock \url{http://dx.doi.org/http://dx.doi.org/10.1063/1.3660809.}

\bibitem[{van~de Meerakker et~al.(2005)van~de Meerakker, Vanhaecke, van~der
  Loo, Groenenboom, and Meijer}]{2005Meerakker-a}
van~de Meerakker S, Vanhaecke N, van~der Loo M, Groenenboom G, Meijer G.
\newblock Direct measurement of the radiative lifetime of vibrationally excited
  {OH} radicals.
\newblock Phys Rev Lett 2005;\hspace{0pt}95:013003.
\newblock \url{http://dx.doi.org/10.1103/PhysRevLett.95.013003.}

\bibitem[{{Carleer}(2001)}]{2001Carleer-a}
{Carleer} MR.
\newblock Wspectra: a {W}indows program to accurately measure the line
  intensities of high-resolution {F}ourier transform spectra.
\newblock In JE~{Russell}, K~{Schaefer}, O~{Lado-Bordowsky}, editors, Remote
  Sensing of Clouds and the Atmosphere V, volume 4168 of \emph{Society of
  Photo-Optical Instrumentation Engineers (SPIE) Conference Series}.
  2001;\hspace{0pt} pages 337--42.
\newblock \url{http://dx.doi.org/10.1117/12.413851.}

\bibitem[{Herman and Wallis(1955)}]{1955Herman-a}
Herman R, Wallis RF.
\newblock Influence of vibration-rotation interaction on line intensities in
  vibration-rotation bands of diatomic molecules.
\newblock J Chem Phys 1955;\hspace{0pt}23(4):637--46.
\newblock \url{http://dx.doi.org/10.1063/1.1742069.}

\bibitem[{Brooke et~al.(2014{\natexlab{a}})Brooke, Ram, Western, Schwenke, Li,
  and Bernath}]{2014Brooke-a}
Brooke JSA, Ram RS, Western CM, Schwenke DW, Li G, Bernath PF.
\newblock Einstein {A} coefficients and oscillator strengths for the
  {A}$^2{\Pi}$-{X}$^2{\Sigma}^+$ (red) and {B}$^2{\Sigma}^+$-{X}$^2{\Sigma}^+$
  (violet) systems and rovibrational transitions in the {X}$^2{\Sigma}^+$ state
  of {CN}.
\newblock {Astrophys J Supp Ser} 2014{\natexlab{a}};\hspace{0pt}210:23.
\newblock \url{http://dx.doi.org/10.1088/0067-0049/210/2/23.}

\bibitem[{Brooke et~al.(2014{\natexlab{b}})Brooke, Bernath, Western, van
  Hemert, and Groenenboom}]{2014Brooke-b}
Brooke JSA, Bernath PF, Western CM, van Hemert MC, Groenenboom GC.
\newblock Line strengths of rovibrational and rotational transitions within the
  {X\TSm} ground state of {NH} 2014{\natexlab{b}};\hspace{0pt}141(5):054310.

\bibitem[{Le~Roy(2004)}]{2004LeRoy-Report-a}
Le~Roy RJ.
\newblock {RKR}1 2.0: A computer program implementing the first-order {RKR}
  method for determining diatomic molecule potential energy functions.
\newblock University of waterloo chemical physics research report, University
  of Waterloo, 2004.
\newblock \url{$<$scienide2.uwaterloo.ca/$\sim$rleroy/rkr/$>$}.

\bibitem[{Le~Roy(2007)}]{2007LeRoy-Report-a}
Le~Roy RJ.
\newblock {LEVEL} 8.0: A computer program for solving the radial
  {S}chr\"{o}dinger equation for bound and quasibound levels.
\newblock University of waterloo chemical physics research report, University
  of Waterloo, 2007.
\newblock \url{$<$scienide2.uwaterloo.ca/$\sim$rleroy/level/$>$}.

\bibitem[{Western(2014)}]{2014Western-Misc-a}
Western CM.
\newblock {PGOPHER}, a program for simulating rotational structure (v.
  7.1.293).
\newblock 2014.
\newblock \url{http://dx.doi.org/10.5523/bris.huflggvpcuc1zvliqed497r2.}

\bibitem[{Bernath(2015)}]{2015Bernath-Book-a}
Bernath PF.
\newblock Spectra of atoms and molecules. 3nd ed.
\newblock Oxford University Press, 2015.

\bibitem[{Brown and Carrington(2003)}]{2003Brown-Book-a}
Brown J, Carrington A.
\newblock Rotational Spectroscopy of Diatomic Molecules.
\newblock Cambridge Molecular Science. Cambridge University Press, 2003.

\bibitem[{{Chackerian} et~al.(1992){Chackerian}, {Goorvitch}, {Benidar},
  {Farrenq}, {Guelachvili}, {Martin}, {Abrams}, and {Davis}}]{1992Chackerian-a}
{Chackerian} C Jr, {Goorvitch} D, {Benidar} A, {Farrenq} R, {Guelachvili} G,
  {Martin} PM, {Abrams} MC, {Davis} SP.
\newblock Rovibrational intensities and electric dipole moment function of the
  {X\DP} hydroxyl radical 1992;\hspace{0pt}48:667--73.
\newblock \url{http://dx.doi.org/10.1016/0022-4073(92)90130-V.}

\bibitem[{{Brown} and {Howard}(1976)}]{1976Brown-a}
{Brown} JM, {Howard} BJ.
\newblock An approach to the anomalous commutation relations of rotational
  angular momenta in molecules 1976;\hspace{0pt}31:1517--25.
\newblock \url{http://dx.doi.org/10.1080/00268977600101191.}

\bibitem[{{Werner} and {Knowles}(1990)}]{1990Werner-a}
{Werner} HJ, {Knowles} P.
\newblock A comparison of variational and nonvariational internally contracted
  multiconfiguration-reference configuration-interaction calculations
  1990;\hspace{0pt}78:175--87.
\newblock \url{http://dx.doi.org/10.1007/BF01112867.}

\bibitem[{{Gdanitz} and {Ahlrichs}(1988)}]{1988Gdanitz-a}
{Gdanitz} RJ, {Ahlrichs} R.
\newblock The averaged coupled-pair functional {(ACPF)}: {A} size-extensive
  modification of {MRCI(SD)} 1988;\hspace{0pt}143:413--20.
\newblock \url{http://dx.doi.org/10.1016/0009-2614(88)87388-3.}

\bibitem[{{Cave} and {Davidson}(1988)}]{1988Cave-a}
{Cave} RJ, {Davidson} ER.
\newblock Quasidegenerate variational perturbation theory and the calculation
  of first-order properties from variational perturbation theory wave functions
  1988;\hspace{0pt}89:6798--814.
\newblock \url{http://dx.doi.org/10.1063/1.455354.}

\bibitem[{{Szalay} and {Bartlett}(1993)}]{1993Szalay-a}
{Szalay} PG, {Bartlett} RJ.
\newblock Multi-reference averaged quadratic coupled-cluster method: a
  size-extensive modification of multi-reference {CI}
  1993;\hspace{0pt}214:481--8.
\newblock \url{http://dx.doi.org/10.1016/0009-2614(93)85670-J.}

\bibitem[{Werner et~al.(2010)Werner, Knowles, Knizia, Manby, {Sch\"{u}tz},
  Celani, Korona, Lindh, Mitrushenkov, Rauhut, Shamasundar, Adler, Amos,
  Bernhardsson, Berning, Cooper, Deegan, Dobbyn, Eckert, Goll, Hampel,
  Hesselmann, Hetzer, Hrenar, Jansen, K\"oppl, Liu, Lloyd, Mata, May,
  McNicholas, Meyer, Mura, Nicklass, O'Neill, Palmieri, Pfl\"uger, Pitzer,
  Reiher, Shiozaki, Stoll, Stone, Tarroni, Thorsteinsson, Wang, and
  Wolf}]{2010Werner-Misc-a}
Werner HJ, Knowles PJ, Knizia G, Manby FR, {Sch\"{u}tz} M, Celani P, Korona T,
  Lindh R, Mitrushenkov A, Rauhut G, Shamasundar KR, Adler TB, Amos RD,
  Bernhardsson A, Berning A, Cooper DL, Deegan MJO, Dobbyn AJ, Eckert F, Goll
  E, Hampel C, Hesselmann A, Hetzer G, Hrenar T, Jansen G, K\"oppl C, Liu Y,
  Lloyd AW, Mata RA, May AJ, McNicholas SJ, Meyer W, Mura ME, Nicklass A,
  O'Neill DP, Palmieri P, Pfl\"uger K, Pitzer R, Reiher M, Shiozaki T, Stoll H,
  Stone AJ, Tarroni R, Thorsteinsson T, Wang M, Wolf A.
\newblock {MOLPRO}, version 2010.1, a package of ab initio programs.
\newblock 2010.
\newblock {h}ttp://www.molpro.net/.

\bibitem[{Werner et~al.(2012)Werner, Knowles, Knizia, Manby, {Sch\"{u}tz},
  Celani, Korona, Lindh, Mitrushenkov, Rauhut, Shamasundar, Adler, Amos,
  Bernhardsson, Berning, Cooper, Deegan, Dobbyn, Eckert, Goll, Hampel,
  Hesselmann, Hetzer, Hrenar, Jansen, K\"oppl, Liu, Lloyd, Mata, May,
  McNicholas, Meyer, Mura, Nicklass, O'Neill, Palmieri, Peng, Pfl\"uger,
  Pitzer, Reiher, Shiozaki, Stoll, Stone, Tarroni, Thorsteinsson, and
  Wang}]{2012Werner-Misc-a}
Werner HJ, Knowles PJ, Knizia G, Manby FR, {Sch\"{u}tz} M, Celani P, Korona T,
  Lindh R, Mitrushenkov A, Rauhut G, Shamasundar KR, Adler TB, Amos RD,
  Bernhardsson A, Berning A, Cooper DL, Deegan MJO, Dobbyn AJ, Eckert F, Goll
  E, Hampel C, Hesselmann A, Hetzer G, Hrenar T, Jansen G, K\"oppl C, Liu Y,
  Lloyd AW, Mata RA, May AJ, McNicholas SJ, Meyer W, Mura ME, Nicklass A,
  O'Neill DP, Palmieri P, Peng D, Pfl\"uger K, Pitzer R, Reiher M, Shiozaki T,
  Stoll H, Stone AJ, Tarroni R, Thorsteinsson T, Wang M.
\newblock {MOLPRO}, version 2012.1, a package of ab initio programs.
\newblock 2012.
\newblock {h}ttp://www.molpro.net/.

\bibitem[{Dunning(1989)}]{1989Dunning-a}
Dunning TH Jr.
\newblock Gaussian basis sets for use in correlated molecular calculations.
  {I}. {T}he atoms boron through neon and hydrogen.
\newblock J Chem Phys 1989;\hspace{0pt}90:1007--23.
\newblock \url{http://dx.doi.org/10.1063/1.456153.}

\bibitem[{Woon and Dunning(1995)}]{1995Woon-a}
Woon DE, Dunning TH Jr.
\newblock Gaussian basis sets for use in correlated molecular calculations.
  {V}. {}core-valence basis sets for boron through neon.
\newblock J Chem Phys 1995;\hspace{0pt}103:4572--85.
\newblock \url{http://dx.doi.org/10.1063/1.470645.}

\bibitem[{Wilson et~al.(1996)Wilson, van Mourik, and Dunning}]{1996Wilson-a}
Wilson AK, van Mourik T, Dunning TH Jr.
\newblock Gaussian basis sets for use in correlated molecular calculations.
  {VI}. {S}extuple zeta correlation consistent basis sets for boron through
  neon.
\newblock J Mol Struct: THEOCHEM 1996;\hspace{0pt}388:339--49.
\newblock \url{http://dx.doi.org/10.1016/S0166-1280(96)80048-0.}

\bibitem[{Knowles and Werner(1985)}]{1985Knowles-a}
Knowles PJ, Werner HJ.
\newblock An efficient second-order {MC SCF} method for long configuration
  expansions.
\newblock Chem Phys Lett 1985;\hspace{0pt}115(3):259--67.
\newblock \url{http://dx.doi.org/10.1016/0009-2614(85)80025-7.}

\bibitem[{Werner and Knowles(1985)}]{1985Werner-a}
Werner HJ, Knowles PJ.
\newblock A second order multiconfiguration {SCF} procedure with optimum
  convergence.
\newblock J Chem Phys 1985;\hspace{0pt}82:5053--63.
\newblock \url{http://dx.doi.org/10.1063/1.448627.}

\bibitem[{{Larsson}(1983)}]{1983Larsson-a}
{Larsson} M.
\newblock Conversion formulas between radiative lifetimes and other dynamical
  variables for spin-allowed electronic transitions in diatomic molecules.
\newblock Astron Astrophys 1983;\hspace{0pt}128:291--8.

\bibitem[{{Laraia} et~al.(2011){Laraia}, {Gamache}, {Lamouroux}, {Gordon}, and
  {Rothman}}]{2011Laraia-a}
{Laraia} AL, {Gamache} RR, {Lamouroux} J, {Gordon} IE, {Rothman} LS.
\newblock Total internal partition sums to support planetary remote sensing.
\newblock Icarus 2011;\hspace{0pt}215:391--400.
\newblock \url{http://dx.doi.org/10.1016/j.icarus.2011.06.004.}

\bibitem[{{Drouin}(2013)}]{2013Drouin-a}
{Drouin} BJ.
\newblock Isotopic spectra of the hydroxyl radical
  2013;\hspace{0pt}117:10076--91.
\newblock \url{http://dx.doi.org/10.1021/jp400923z.}

\bibitem[{{Mel{\'e}ndez}(2004)}]{mel04}
{Mel{\'e}ndez} J.
\newblock A low solar oxygen abundance from the first-overtone {OH} lines
  2004;\hspace{0pt}615:1042--7.
\newblock \url{http://dx.doi.org/10.1086/424591.}

\bibitem[{{Kaeufl} et~al.(2004){Kaeufl}, {Ballester}, {Biereichel}, {Delabre},
  {Donaldson}, {Dorn}, {Fedrigo}, {Finger}, {Fischer}, {Franza}, {Gojak},
  {Huster}, {Jung}, {Lizon}, {Mehrgan}, {Meyer}, {Moorwood}, {Pirard},
  {Paufique}, {Pozna}, {Siebenmorgen}, {Silber}, {Stegmeier}, and
  {Wegerer}}]{kae04}
{Kaeufl} HU, {Ballester} P, {Biereichel} P, {Delabre} B, {Donaldson} R, {Dorn}
  R, {Fedrigo} E, {Finger} G, {Fischer} G, {Franza} F, {Gojak} D, {Huster} G,
  {Jung} Y, {Lizon} JL, {Mehrgan} L, {Meyer} M, {Moorwood} A, {Pirard} JF,
  {Paufique} J, {Pozna} E, {Siebenmorgen} R, {Silber} A, {Stegmeier} J,
  {Wegerer} S.
\newblock {CRIRES}: a high-resolution infrared spectrograph for {ESO}'s {VLT}.
\newblock In AFM {Moorwood}, M~{Iye}, editors, Ground-based Instrumentation for
  Astronomy, volume 5492 of \emph{Society of Photo-Optical Instrumentation
  Engineers (SPIE) Conference Series}. 2004;\hspace{0pt} pages 1218--27.
\newblock \url{http://dx.doi.org/10.1117/12.551480.}

\bibitem[{{Wilson} et~al.(2010){Wilson}, {Hearty}, {Skrutskie}, {Majewski},
  {Schiavon}, {Eisenstein}, {Gunn}, {Blank}, {Henderson}, {Smee}, {Barkhouser},
  {Harding}, {Fitzgerald}, {Stolberg}, {Arns}, {Nelson}, {Brunner}, {Burton},
  {Walker}, {Lam}, {Maseman}, {Barr}, {Leger}, {Carey}, {MacDonald}, {Horne},
  {Young}, {Rieke}, {Rieke}, {O'Brien}, {Hope}, {Krakula}, {Crane}, {Zhao},
  {Carr}, {Harrison}, {Stoll}, {Vernieri}, {Holtzman}, {Shetrone},
  {Allende-Prieto}, {Johnson}, {Frinchaboy}, {Zasowski}, {Bizyaev},
  {Gillespie}, and {Weinberg}}]{wil10}
{Wilson} JC, {Hearty} F, {Skrutskie} MF, {Majewski} S, {Schiavon} R,
  {Eisenstein} D, {Gunn} J, {Blank} B, {Henderson} C, {Smee} S, {Barkhouser} R,
  {Harding} A, {Fitzgerald} G, {Stolberg} T, {Arns} J, {Nelson} M, {Brunner} S,
  {Burton} A, {Walker} E, {Lam} C, {Maseman} P, {Barr} J, {Leger} F, {Carey} L,
  {MacDonald} N, {Horne} T, {Young} E, {Rieke} G, {Rieke} M, {O'Brien} T,
  {Hope} S, {Krakula} J, {Crane} J, {Zhao} B, {Carr} M, {Harrison} C, {Stoll}
  R, {Vernieri} MA, {Holtzman} J, {Shetrone} M, {Allende-Prieto} C, {Johnson}
  J, {Frinchaboy} P, {Zasowski} G, {Bizyaev} D, {Gillespie} B, {Weinberg} D.
\newblock {T}he {A}pache {P}oint {O}bservatory {G}alactic {E}volution
  {E}xperiment {(APOGEE)} high-resolution near-infrared multi-object fiber
  spectrograph.
\newblock In Society of Photo-Optical Instrumentation Engineers (SPIE)
  Conference Series, volume 7735 of \emph{Society of Photo-Optical
  Instrumentation Engineers (SPIE) Conference Series}. 2010;\hspace{0pt}
  page~1.
\newblock \url{http://dx.doi.org/10.1117/12.856708.}

\bibitem[{{Park} et~al.(2014){Park}, {Jaffe}, {Yuk}, {Chun}, {Pak}, {Kim},
  {Pavel}, {Lee}, {Oh}, {Jeong}, {Sim}, {Lee}, {Nguyen Le}, {Strubhar},
  {Gully-Santiago}, {Oh}, {Cha}, {Moon}, {Park}, {Brooks}, {Ko}, {Han}, {Nah},
  {Hill}, {Lee}, {Barnes}, {Yu}, {Kaplan}, {Mace}, {Kim}, {Lee}, {Hwang}, and
  {Park}}]{par14}
{Park} C, {Jaffe} DT, {Yuk} IS, {Chun} MY, {Pak} S, {Kim} KM, {Pavel} M, {Lee}
  H, {Oh} H, {Jeong} U, {Sim} CK, {Lee} HI, {Nguyen Le} HA, {Strubhar} J,
  {Gully-Santiago} M, {Oh} JS, {Cha} SM, {Moon} B, {Park} K, {Brooks} C, {Ko}
  K, {Han} JY, {Nah} J, {Hill} PC, {Lee} S, {Barnes} S, {Yu} YS, {Kaplan} K,
  {Mace} G, {Kim} H, {Lee} JJ, {Hwang} N, {Park} BG.
\newblock Design and early performance of {IGRINS} ({I}mmersion {G}rating
  {I}nfrared {S}pectrometer).
\newblock In Society of Photo-Optical Instrumentation Engineers (SPIE)
  Conference Series, volume 9147 of \emph{Society of Photo-Optical
  Instrumentation Engineers (SPIE) Conference Series}. 2014;\hspace{0pt}
  page~1.
\newblock \url{http://dx.doi.org/10.1117/12.2056431.}

\bibitem[{{Asplund} et~al.(2004){Asplund}, {Grevesse}, {Sauval}, {Allende
  Prieto}, and {Kiselman}}]{asp04}
{Asplund} M, {Grevesse} N, {Sauval} AJ, {Allende Prieto} C, {Kiselman} D.
\newblock Line formation in solar granulation. {IV. [O I], O I} and {OH} lines
  and the photospheric {O} abundance 2004;\hspace{0pt}417:751--68.
\newblock \url{http://dx.doi.org/10.1051/0004-6361:20034328.}

\bibitem[{{Hinkle} et~al.(2013){Hinkle}, {Wallace}, {Ram}, {Bernath}, {Sneden},
  and {Lucatello}}]{hin13}
{Hinkle} KH, {Wallace} L, {Ram} RS, {Bernath} PF, {Sneden} C, {Lucatello} S.
\newblock The magnesium isotopologues of {MgH} in the
  {A}$^{2}${$\Pi$}-{X}$^{2}${$\Sigma$}$^{+}$ system 2013;\hspace{0pt}207:26.
\newblock \url{http://dx.doi.org/10.1088/0067-0049/207/2/26.}

\bibitem[{{Ram} et~al.(2014){Ram}, {Brooke}, {Bernath}, {Sneden}, and
  {Lucatello}}]{ram14}
{Ram} RS, {Brooke} JSA, {Bernath} PF, {Sneden} C, {Lucatello} S.
\newblock Improved line data for the {S}wan system $^{12}${C}$^{13}${C}
  isotopologue 2014;\hspace{0pt}211:5.
\newblock \url{http://dx.doi.org/10.1088/0067-0049/211/1/5.}

\bibitem[{{Sneden} et~al.(2014){Sneden}, {Lucatello}, {Ram}, {Brooke}, and
  {Bernath}}]{sne14}
{Sneden} C, {Lucatello} S, {Ram} RS, {Brooke} JSA, {Bernath} P.
\newblock Line lists for the {A}$^{2}${$\Pi$}-{X}$^{2}${$\Sigma$}$^{+}$ (red)
  and {B}$^{2}${$\Sigma$}$^{+}$-{X}$^{2}${$\Sigma$}$^{+}$ (violet) systems of
  {CN}, $^{13}${C}$^{14}${N}, and $^{12}${C}$^{15}${N}, and application to
  astronomical spectra 2014;\hspace{0pt}214:26.
\newblock \url{http://dx.doi.org/10.1088/0067-0049/214/2/26.}

\bibitem[{{Sneden}(1973)}]{sne73}
{Sneden} C.
\newblock The nitrogen abundance of the very metal-poor star {HD} 122563.
  1973;\hspace{0pt}184:839--49.
\newblock \url{http://dx.doi.org/10.1086/152374.}

\bibitem[{{Kurucz}(2011)}]{kur11}
{Kurucz} RL.
\newblock Including all the lines 2011;\hspace{0pt}89:417--28.
\newblock \url{http://dx.doi.org/10.1139/p10-104.}

\bibitem[{{Lawler} et~al.(2014){Lawler}, {Wood}, {Den Hartog}, {Feigenson},
  {Sneden}, and {Cowan}}]{law14}
{Lawler} JE, {Wood} MP, {Den Hartog} EA, {Feigenson} T, {Sneden} C, {Cowan} JJ.
\newblock Improved {V I} log(gf) values and abundance determinations in the
  photospheres of the {S}un and metal-poor star {HD} 84937
  2014;\hspace{0pt}215:20.
\newblock \url{http://dx.doi.org/10.1088/0067-0049/215/2/20.}

\bibitem[{{Sneden} et~al.(2009){Sneden}, {Lawler}, {Cowan}, {Ivans}, and {Den
  Hartog}}]{sne09}
{Sneden} C, {Lawler} JE, {Cowan} JJ, {Ivans} II, {Den Hartog} EA.
\newblock New rare earth element abundance distributions for the sun and five
  r-process-rich very metal-poor stars 2009;\hspace{0pt}182:80--96.
\newblock \url{http://dx.doi.org/10.1088/0067-0049/182/1/80.}

\bibitem[{{Asplund} et~al.(2009){Asplund}, {Grevesse}, {Sauval}, and
  {Scott}}]{asp09}
{Asplund} M, {Grevesse} N, {Sauval} AJ, {Scott} P.
\newblock The chemical composition of the {S}un 2009;\hspace{0pt}47:481--522.
\newblock \url{http://dx.doi.org/10.1146/annurev.astro.46.060407.145222.}

\bibitem[{{Holweger} and {M{\" u}ller}(1974)}]{hol74}
{Holweger} H, {M{\" u}ller} EA.
\newblock The photospheric barium spectrum - solar abundance and collision
  broadening of {BA II} lines by hydrogen.
\newblock Sol Phys 1974;\hspace{0pt}39:19--30.
\newblock \url{http://dx.doi.org/10.1007/BF00154968.}

\bibitem[{{Hinkle} et~al.(2000){Hinkle}, {Wallace}, {Valenti}, and
  {Harmer}}]{hin00}
{Hinkle} K, {Wallace} L, {Valenti} J, {Harmer} D.
\newblock Visible and Near Infrared Atlas of the {A}rcturus Spectrum 3727-9300
  {$\AA$}.
\newblock Astronomical Society of the Pacific Monograph Publications, San
  Francisco, 2000.

\bibitem[{{Allende Prieto} et~al.(2008){Allende Prieto}, {Majewski},
  {Schiavon}, {Cunha}, {Frinchaboy}, {Holtzman}, {Johnston}, {Shetrone},
  {Skrutskie}, {Smith}, and {Wilson}}]{all08}
{Allende Prieto} C, {Majewski} SR, {Schiavon} R, {Cunha} K, {Frinchaboy} P,
  {Holtzman} J, {Johnston} K, {Shetrone} M, {Skrutskie} M, {Smith} V, {Wilson}
  J.
\newblock {APOGEE}: The {A}pache {P}oint {O}bservatory {G}alactic {E}volution
  {E}xperiment.
\newblock Astronomische Nachrichten 2008;\hspace{0pt}329:1018.
\newblock \url{http://dx.doi.org/10.1002/asna.200811080.}

\bibitem[{{Ram{\'{\i}}rez} and {Allende Prieto}(2011)}]{2011Ramirez}
{Ram{\'{\i}}rez} I, {Allende Prieto} C.
\newblock Fundamental parameters and chemical composition of {A}rcturus
  2011;\hspace{0pt}743:135.
\newblock \url{http://dx.doi.org/10.1088/0004-637X/743/2/135.}

\end{thebibliography}

\end{document}